\documentclass[twocolumn,showpacs,cite]{revtex4}
\usepackage{epsfig,pslatex,latexsym,bm,revsymb,times,amssymb,amsmath,graphicx}
\pdfoutput=1
\begin{document}
\author{M. Prada} 
\email{prada@wisc.edu}
\affiliation{Instituto de Ciencias Materiales de Madrid, ICMM-CSIC, Sor Juana Ines de la Cruz 3, Madrid, Spain}
\affiliation{Physics Department, University of Wisconsin-Madison, 1150 University Ave., Madison, Wisconsin 53705, USA}
\affiliation{I. Institut f\"ur Theoretische Physik, Universit\"at Hamburg, Jungiusstr. 9, 20355 Hamburg, Germany}
\author{G. Platero}
\affiliation{Instituto de Ciencias Materiales de Madrid, ICMM-CSIC, Sor Juana Ines de la Cruz 3, Madrid, Spain}
\date{\today}
\title{Double coupled electron shuttle} 
\newcommand{\ud}{\mathrm{d}}
\newcommand{\ue}{\mathrm{e}}
%
\begin{abstract}
A nano-shuttle consisting of two movable islands connected in series and 
integrated between two contacts is studied. 
We evaluate the electron transport through the system in the presence of a source-drain voltage 
with and without an rf excitation. 
We evaluate the response of the system in terms of the net 
direct current enhanced by the mechanical motion of the oscillators.
An introduction to the charge stability diagram is given in terms of electrochemical 
potentials and mechanical displacements.  
The low capacitance of the islands allows the observation of Coulomb blockade even at room temperature. 
Using radio frequency excitations, the nonlinear dynamics of the system is studied. 
The oscillators can be tuned to unstable regions where mechanically assisted transfer of 
electrons can further increase the amplitude of motion, resulting of a net energy being pumped into the system. 
The resulting amplified response can be exploited to design a mechanical motion detector 
of nanoscale objects. 
\end{abstract}
\maketitle
\section{Introduction}
Recent experiments on coupled shuttles show intriguing effects 
arising from the coupling of the electrostatics and mechanical degrees of freedom \cite{scheible_apl04,ck}. 
The  imprints of single-electron effects have been observed in these systems, such as Coulomb
blockade \cite{ck2,cohen} and gate-voltage-dependent oscillations of the conductance \cite{ck3}.
The increasing relevance of nano electromechanical system (NEMS) is a result of their potential industrial applications  
\cite{beeby_review,li_n}. 
NEMS offer the possibility to realize, for instance, nanomechanical switches \cite{subramanian_acs}, circuits
\cite{blicknjop,mahboob,roukesNMC}, 
electronic transducers \cite{roukes_nano,bartsch_acs, oroszlany_acs}, solar cells \cite{solar_cells} 
or high-sensitive charge \cite{scheible_apl04}, spin \cite{rugarNature} and
mass sensors \cite{rugarPRL,roukesMD}, as well as the general study of
nonlinear dynamics of oscillators and resonators \cite{pistolesi, ahn, haupt, merlo}.
In particular, there is a growing interest in parametrically driven nonlinear 
systems \cite{villanueva_nano,  midvedt_nano}, where  a smooth change in the value 
of a parameter results in a sudden change of the response of a system. 

Our aim is to present here a theoretical study on the electromechanics of a 
coupled shuttle with an external electrical excitation.
The ``smoothly'' changing parameters are the intensity and the frequency of the excitation, 
and the response is the observable direct current through the system.  
In this context, we consider mechanically assisted electron transport through a system of 
movable low-capacitance nanoislands connected in series between two electrodes.
The mechanical motion of the islands changes the mutual capacitance of the system and
the tunneling processes, hence affecting the current through the system.
We can thus say that the mechanical and electronic degrees of freedom are coupled.
An rf excitation allows us to study the effects of nonlinear mechanics,
where multiple stability can be achieved by tuning the frequency and intensity of the
excitation in the coupled mode regime.
In the unstable regions the response of the system is greatly amplified, suggesting a practical
scheme for detection of instabilities in the mechanical motion of nanoscale objects. 

\begin{figure}[!hbt]
\centering\includegraphics[angle=0, width = 0.450\textwidth]{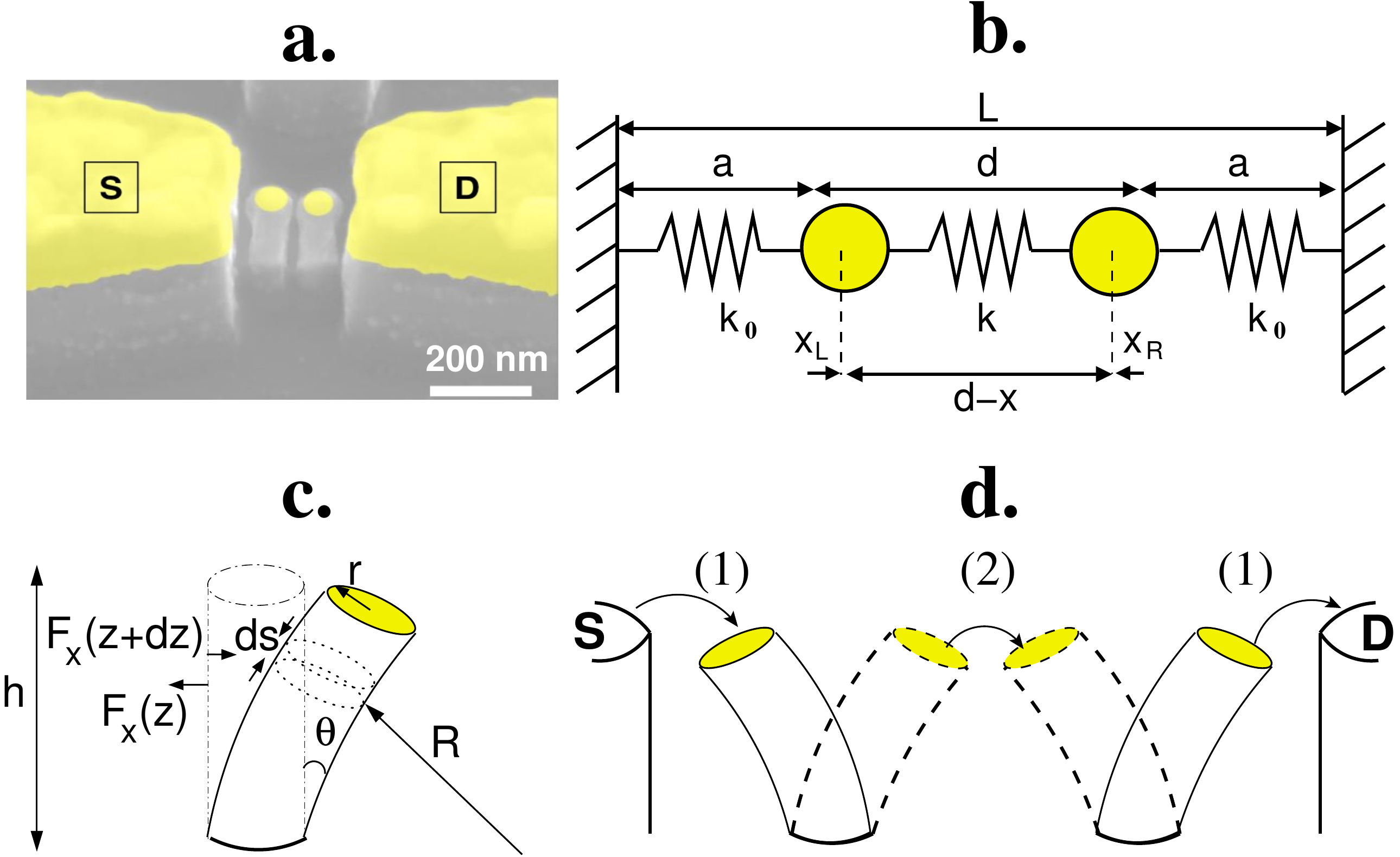}
\caption{\it \footnotesize (a) SEM image of a coupled shuttle consisting 
of two Si-based nanopillars,  
(b) double harmonic oscillator, with coupling spring constants $k$ and $k_0$, and 
mechanical displacements $x_L$ and $x_R$, and
(c) forces that generate the displacement $x$ in a differential element $\ud s$ of the nanopillar.
(d) The flexural mode where the center of mass is at rest. 
}
\label{fig1abc}
\end{figure}

An example of the device that we analyze is the one developed by Kim {\it et al.} \cite{ck2}, 
consisting of a double pillar structure with a gold nanoisland on top. 
[see the SEM image of Fig. \ref{fig1abc}(a)]. 
The device can be described in terms of two sets of characteristic quantities. 
The first one is the relative displacement of the islands, 
$x = x_L - x_R = r_0\cos{\omega t}$ [see Fig. \ref{fig1abc}(b)], with $r_0$ being the amplitude of the oscillations
and $\omega$ the vibration frequency. 
The mutual capacitances and resistances of the device depend on this quantity, 
affecting as well the other important set of parameters: 
the electrostatic free energy for a given charge configuration,  $F(m_i)$,
where $m_i$ labels the charge state of the device. 

This work is organized as follows:
in Sec. \ref{mech}, we describe the purely mechanical aspects of the nanopillars in a flexural mode.
Sec. \ref{CED} then describes the electrostatics of the system, introducing the free energy and 
chemical potentials as a function of the mechanical displacements. 
In Sec. \ref{coupled} we consider the coupling of the mechanical and electronic degrees of freedom. 
A dynamic equation is derived, and we conclude with a master equation to describe electron transfer 
processes between the contacts. 
We evaluate in detail the small oscillations limit and the shuttling regime within the Coulomb 
blockade limit. 
We also give an expression for the dissipated and absorbed power by the device.
Finally, we devote Sec. \ref{conclusions} to conclusions. 

\section{Mechanical aspects of the coupled oscillator} 
\label{mech}
The system of our interest is represented in Fig. \ref{fig1abc}.  
Two nanopillars of height $h$ are operated in a flexural mode where the center of mass 
remains at rest in most cases, mechanically assisting electronic transport across the system. 
We consider first a single pillar, as in Fig. \ref{fig1abc} (c). 
The beam can vibrate along $x$, with displacements $x(z,t)$. 
A differential element of length $dz$ and cross-sectional area $A$
is subject to forces $F_x(z+dz)$ and $-F_x(z)$ on each face, directed along $x$, and
torques $M_y(z+dz)$ and $-M_y(z)$, directed along $y$. Balancing linear forces and
imposing that there is no net torque \cite{andrewsbook}, we have 
\begin{eqnarray}
&&F_x(z+dz)-F_x(z) = \rho A {{d}}z \frac{\partial^2x}{\partial t^2},  
\nonumber \\
&&F_x(z+dz)dz +  M_y(z+dz) - M_y(z) = 0  , 
\end{eqnarray}
where $M_y = EI_x(\partial^2x/\partial z^2)$, $E$ being the Young modulus and $I_x= \pi r^4/4$, the 
second moment of area of a cylinder.  
Expanding about the point $z$ and keeping only first-order terms in d$z$, 
we find the Euler-Bernoulli equation (note that we are neglecting the damping force for now):
\[ 
E I_x\frac{\partial^4x}{\partial z^4} = -\rho A \frac{\partial^2x}{\partial t^2} , 
\]
with solutions
$ x_n(z,t) = [a_n(\cos{\beta_nz} -\cosh{\beta_nz} ) + b_n(\sin{\beta_nz} -\sinh{\beta_nz})]\cos{\omega_nt}$.
With the boundary conditions $x(0,t) = \partial_z x(0,t) = 0$
and $\partial^2_z x(h,t) = \partial^3_z x(h,t) = 0$, we find numerically
$\beta_n h$ = 1.875, 4.694, 7.855, $\dots$, with $\omega_n = \sqrt{(EI_x/\rho A)}\beta_n^2$ 
and $a_n/b_n = -1.362, =-0.982, =-1.008, =-1.000, \dots$ \cite{andrewsbook}.  
For the first mode, we can estimate that $\omega_0 \simeq$ 240 MHz for a typical nanopillar
with a radius $r\sim$ 30 nm and using a Young modulus of $E$ = 150 GPa \cite{hopcroft}.
At the metallic islands situated on top of the pillars ($z=h$), the mechanical movement thus can be described 
as a harmonic oscillator,  
$m \ddot x_i + \omega_0^2 x_i = 0$.

We now consider the double oscillator in its coupled mode, as depicted in Fig. \ref{fig1abc}(b). 
Two harmonic oscillators of mass $m$ and spring constants $k_0 = m\omega_0^2$, with 
coupling spring constant $k =\varsigma m \omega_0^2$, where $\varsigma$ 
is a parameter that quantifies the coupling of the oscillators.
The dynamics of the system can be derived from its Lagrangian,
\begin{equation}
\mathcal{L} = \frac{1}{2}m\dot{x}_L^2 + \frac{1}{2}m\dot{x}_R^2 -
\frac{1}{2}m\omega_0^2\left[ x_L^2 + x_R^2+\varsigma(x_R-x_L)^2\right].
\label{eq:lamda}
\end{equation}
The problem suggests using the new coordinates, where $X = (x_L+x_R)/2$ is the center of mass and
$x = x_L-x_R$, the relative displacement, giving  
 \[
x(t)  = r_0 \cos{(\omega t + \varphi)}, \quad 
X(t)  = X_0 \cos{(\omega_0 t + \varphi^\prime)},
\]
where $r_0$, $X_0$, $\varphi$ and $\varphi^\prime$ are determined by the initial conditions. 
We see that the center of mass moves with the natural frequency $\omega_0$, 
whereas the relative coordinate moves with a higher frequency, 
$\omega = (1+2\varsigma)\sqrt{k_0/m}$ = $\sqrt{k'/m}$, $k' = k_0(1+2\varsigma)^2$.
Two limiting cases can be considered: strong coupling (SC), where $\varsigma\gg 1$, 
and weak coupling (WC), with $\varsigma \ll 1$. In the SC regime, the movement of one
oscillator is quickly transferred to the second, whereas in the WC regime, the movement of the
first oscillator is slowly transferred to the second one, which is the situation we expect to 
encounter in our system. 

\section{Electrodynamics of two metallic movable grains between two contacts}
\label{CED}
\subsection{Free energy of movable coupled metallic grains}
\begin{figure}[!hbt]
\centering\includegraphics[angle=0, width = 0.350\textwidth]{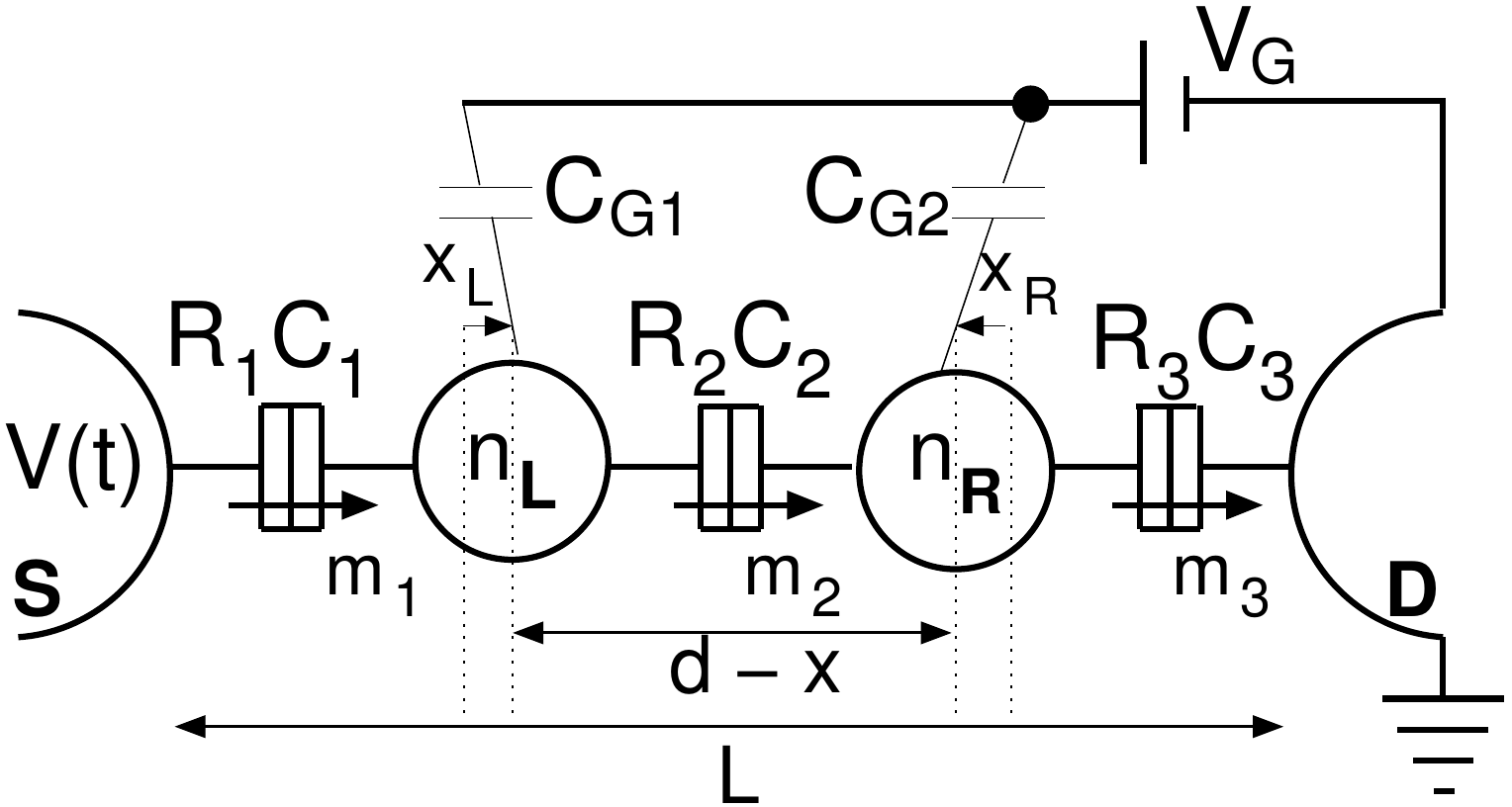}
\caption{\it \footnotesize A schematic picture of the double-island structure  
with the voltage sources and various capacities in the system. 
The two circles denote the islands with $n_{L,R}$ excess electrons. 
The distance between the islands is given in terms of their relative displacement, $x(t)$. 
A bias $V(t)$ is applied to the left contact while maintaining the right one grounded. 
}
\label{fig1}
\end{figure}
The circuit diagram of  our system is depicted in Fig. \ref{fig1}: 
two oscillating, capacitively coupled metallic islands with $n_L$ ($n_R$)  
excess electrons in the left (right) island, resulting in a sequence of 
three tunnel junctions, $i$ = 1, 2, 3, each characterized by a 
resistance and a capacitance, $R_i$ and $C_i$. 
$n_{L,R}$ is determined by the charge accumulation in the junctions, $m_i$ 
\begin{eqnarray}
\label{eq:ns}
n_L&=&m_1-m_2,\nonumber \\
n_R&=&m_2-m_3.  
\end{eqnarray}
A bias $V(t)$ is applied to the left contact, while maintaining the right one 
grounded.

The capacitances of the (disk shaped) left and right island $C_{L,R}$ 
and their mutual capacitance $C_2$  can be expressed in terms of 
their radii, $r_{L,R}$, 
\begin{equation}
C_{L,R} \simeq 8\epsilon r_{L,R}; \qquad C_2\simeq  8\epsilon \frac{r_Lr_R}{(d-x)} = C_2^0\frac{1}{1-x/d},
\label{eq:caps}
\end{equation}
where 
$d$ is the equilibrium distance between the islands. 
Here, $\epsilon$ is the dielectric constant of the material surrounding the 
electrodes,  $\epsilon =  \epsilon_0\epsilon_r$. In our case, $\epsilon_r\sim$ 1 
(air), so the dielectric constant is close to the vacuum one, $\epsilon_0$.

A full derivation of the free energy is given in Appendix \ref{app:fe}. 
The free energy in the linear transport regime [{\it i.e.}, for $V(t)\simeq 0$] 
in terms of the $m_i$ and $n_\alpha$ given by Eq. (\ref{eq:ns})
reads  
\begin{eqnarray}
F(\{n_\alpha,m_i\}) & =& \frac{1}{2}n_L^2E_{CL} + \frac{1}{2}n_R^2 E_{CR} + n_Ln_RE_{CC} - W(\{m_i\}); \nonumber\\
W(\{m_i\}) &=& \frac{1}{|e|} \left\{
m_1[V_G(C_{G1}E_{CL} + C_2^0E_{CC}) ] + 
\right.\nonumber \\  &+&\left.
m_2[V_G(C_{G1}(E_{CL}-E_{CC}) -C_{G2}(E_{CR}-E_{CC})]
\right.\nonumber \\ &+& \left.
m_3[ -V_G(C_{G1}E_{CC} + C_{G2}E_{CR})] 
\right\},
\label{eq:F}
\end{eqnarray}
where $E_{CL(CR)}$ is the charging energy of the left (right) island and $E_{CC}$ is 
the electrostatic coupling energy, which denotes the change in energy of one island when 
an electron is added to the other island. These energies can be expressed in terms of $x$, 
using Eq. (\ref{eq:caps}),
\begin{equation}
\label{Eq:Eci}
E_{CL/CR} = \frac{e^2}{C_{L/R}} 
\left( \frac{1}{1-\frac{r_Lr_R}{(d-x)^2} }
\right);\ 
E_{CC} = \frac{e^2}{C_2^0} 
\left( \frac{1}{\frac{(d-x)^2}{r_Lr_R}-1 }
\right).
\end{equation}
We note that the charging energy of the individual islands do not change dramatically with 
the oscillations, since $(d-x)\ge r_L + r_R$ implies (using $r_L\sim r_R = r$) 
$ E_{CL/CR}^0< E_{CL/CR}< 4  E_{CL/CR}^0/3$, with $ E_{CL/CR}^0 = e^2/8\epsilon r_{L/R}$. 
On the other hand, the coupling energy in the regime of strong oscillations can take 
different limits: 
When the islands are far apart, $x<0$ and $(d-x)^2 \gg r_L r_R$, then $E_{CC}\to 0$ and the free energy 
given in  (\ref{eq:F}) is formally equivalent to the sum of the energies of two independent islands, 
\[
F\simeq \frac{(V_GC_{G1}-n_L e )^2}{2e^2}E_{CL} + \frac{(V_GC_{G2}-n_R e )^2}{2e^2}E_{CR}  + f(\{m_i\},V_G^2), 
\] 
with $f$ being a function that does not depend on $x$. 
As they approach, $E_{CC}$ becomes larger, with $E_{CC}\lesssim e^2/12\epsilon r$. 
For $(d-x)\sim 2r$ and $C_{G1}\sim C_{G2} = C_G$, we separate the $x$--dependent and $x$--independent ($g$) terms, 
\[
F\simeq \frac{(4V_GC_{G}-(n_L+n_R) e )^2}{2e^2}E_{CC} + g(\{m_i\},V_G^2),  
\] 
which corresponds to the energy of a single island with charge $n_L + n_R$. 
Hence, in the large amplitude of oscillations limit, the device oscillates between the 
two independent islands regime and a single large island regime, giving rise to a rich 
structure in the response. 
\subsection{Transport in the Coulomb blockade limit}
\label{transport}
\begin{figure}[!hbt]
\centering\includegraphics[angle=0, width = 0.450\textwidth]{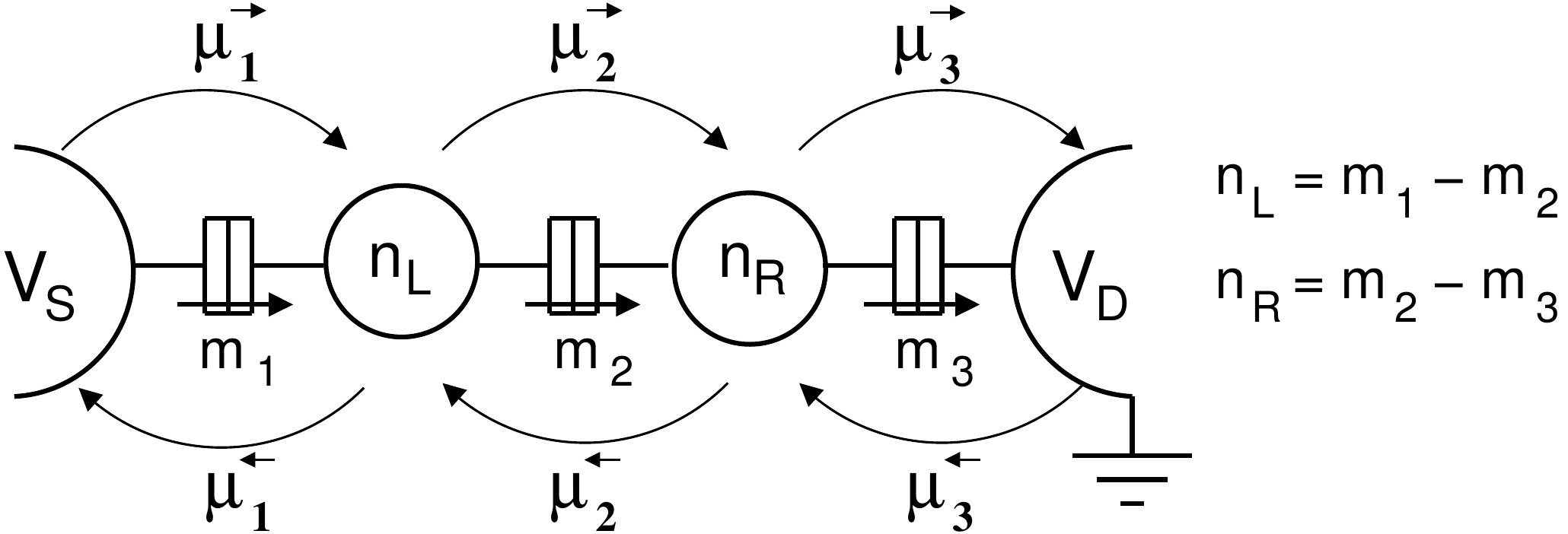}
\caption{\it \footnotesize
Schematic representation of the tunneling processes in
a system of two islands attached to the S and D leads.
Six different processes are given in terms of chemical potentials,
$\mu^\rightleftarrows_i$, $i$ = 1, 2, 3, as indicated in the picture.
}
\label{figmus}
\end{figure}

We consider the electric transport within the classical regime studied by Kulik and Shekhter 
\cite{kulik75}. 
Most of the single-electron effects can be explained in terms of lowest-order perturbation theory, 
since higher-order tunneling processes, as cotunneling, are exponentially suppressed 
due to the mechanical motion of the islands. 
The charge state is given in terms of the probabilities of having $n_L$ excess electrons 
in the left island and $n_R$ excess electrons in the right island, $P_{n_L,n_R}$. 
The tunneling processes are described in terms of transition rates within the ``Orthodox'' model
\cite{averin}, resulting in an equation of motion that describes the evolution of the 
charge with time. 
In such a picture, $\overleftrightarrow\Gamma_{n_L,n_R}^j$ denotes the tunneling rate 
across junction $j$ in the forward or backward direction, having $n_L$ and $n_R$ excess 
electrons in either island: 
\begin{equation}
\label{eqGamma}
\overleftrightarrow\Gamma^{i}_{n_L,n_R} = 
\frac{-\mu_{i}^\rightleftarrows}
     {(1-\mathrm{e}^{\mu_i^\rightleftarrows/k_BT})e^2R_i}.
\end{equation}
Here, $R_i$ is the resistance in the $i$-th junction, which depends exponentially on the 
displacement of the islands. On the flexural mode, we have 
\begin{equation}
R_{1,3} = R_{1,3}^0e^{\frac{x}{2\lambda}}; \quad R_2 = R_{2}^0e^{-\frac{x}{\lambda}}
\label{eq:resis}
\end{equation}
where $R_i^0$ is the static resistance on junction $i$ and $\lambda$ is the phenomenological tunneling length. 
The electrochemical potentials or addition energies,  $\mu^\rightleftarrows_i$, 
denote the  energy an electron needs to overcome in order to tunnel across the junction $i$ 
while keeping fixed the number of electrons in junction $j$ ($j\neq i$), 
$\mu^\rightleftarrows_i (n_L,n_R) = F(m_i\pm 1) - F(m_i) $.  
Figure \ref{figmus} shows schematically the six different processes across the three 
junctions.  
Using the expression for the free energy (\ref{eq:F}), the electrochemical 
potentials for the six possible processes of Fig. \ref{figmus} are  
\begin{widetext}
\begin{eqnarray}
\mu_{1}^\rightleftarrows (n_L,n_R)&=& 
\left[
\left(\frac{1}{2}\pm n_L\right)\pm
\delta_L n_R\mp
\frac{V_G}{|e|} \left(C_{G1}+C_{G2}\delta_L \right) \pm 
\frac{V(t)}{|e|}\left((C_L-C_1) -\delta_L C_2^0\right)
\right]E_{CL}
\nonumber\\
\mu_{2}^\rightleftarrows(n_L,n_R)&=&  
\left[ 1\mp n_L \mp n_R \mp\frac{V(t)}{|e|}C_1 \pm
\frac{V_G}{|e|}(C_{G1}-C_{G2})
\right]
(E_{CL}-E_{CC}) 
\nonumber\\
\mu_{3}^\rightleftarrows (n_L,n_R)&=&  
\left[
\left(\frac{1}{2}\mp n_R\right)\mp \delta_Rn_L\mp
\frac{V_G}{|e|} \left(C_{G2}+C_{G1}\delta_R\right)\pm
\frac{V(t)}{|e|} C_1\delta_R 
\right]E_{CR} ,
\label{eq:mus2}
\end{eqnarray}
\end{widetext}
where we have defined $\delta_{L/R} = r_{L/R}/(d-x)$ and made the approximation $r_L \simeq r_R$ in
the second equation. 
The master equation approach extended to multiple junctions reads
\begin{widetext}
\begin{eqnarray}
\dot P_{n_L,n_R} &=& \overrightarrow\Gamma^1_{n_L-1,n_R} P_{n_L-1,n_R} +\overleftarrow\Gamma^1_{n_L+1,n_R} P_{n_L+1,n_R} +
\overrightarrow\Gamma^2_{n_L+1,n_R-1} P_{n_L+1,n_R-1} + 
\overleftarrow\Gamma^2_{n_L-1,n_R+1} P_{n_L-1,n_R+1} + 
\overrightarrow\Gamma^3_{n_L,n_R+1} P_{n_L,n_R+1} + 
\nonumber \\ &+&
\overleftarrow\Gamma^3_{n_L,n_R-1} P_{n_L,n_R-1} - 
\sum_{j,\rightleftarrows}\overleftrightarrow\Gamma^j_{n_L,n_R} P_{n_L,n_R}.
\label{eq:master1}
\end{eqnarray}
\end{widetext}
In order to understand qualitatively the charge transport in this mechanically movable 
device, let us focus on the charge transfer from the source to the left island 
in the absence of a gate voltage and in the neutral charge state, $n_L = n_R =0$. 
We can express the corresponding chemical potential in terms of $x$ by
using (\ref{eq:caps}) in (\ref{eq:mus2}): 
\[
\mu_1^\rightarrow \simeq \frac{e^2(1-2x/d)}{2C_L(1+\delta_L\delta_R)} \left[ 
1 + \frac{V_S}{|V_{\mathrm{th}}^{01}|(1-x/d)} 
\right] , 
\] 
where we have defined the static voltage threshold $|V_{\mathrm{th}}^{01}|$ = $|e|/2|(C_L-C_1 - C_2^0\delta_L)|$ 
$\sim |e|/2[C_2^0(1-\delta_L)]$ 
(note that the junction capacitances in the contacts are $C_1$ = $C_L$ - $C_2$ - $C_{G1}$ and 
$C_3$ = $C_R$ - $C_2$ - $C_{G2}$). 
At zero temperature, a tunneling event requires a negative chemical potential $\mu_1^\rightarrow<0$, 
involving, in the absence of mechanical movement, a negative source voltage $V_S<0$. 
For a movable system, the inequality for a tunneling event to occur reads 
\[
|V_S| >  |V_{\mathrm{th}}^{01}| (1-x/d). 
\]
If the voltage is in magnitude just below the threshold, say $|V_S| = |V_{\mathrm{th}}^{01}|(1-\Delta_\epsilon)$  
with $0<\Delta_\epsilon<1$, 
then a negative chemical potential requires $(1-\Delta_\epsilon)$ $>$ $(1-x/d)$, or
$x/d>\Delta_\epsilon >0$, involving the island 
separating from the left contact [recall that $x$ is {\emph{positive}} when both islands {\emph{separate}} from the contacts, as 
in step (2) of Fig. \ref{fig1abc}(d)]. 
However, the resistance increases exponentially with the 
distance, hence suppressing the tunneling process. Small oscillations are possible due to the 
elastic force, but the island will not be pushed back and forth by Coulomb forces.   
On the contrary, beyond the threshold, say 
$|V_S| = V_{\mathrm{th}}^{01}(1+\Delta_\epsilon)$, a negative potential no longer requires negative $x$. 
Tunneling of one excess electron as the island approaches the contact then becomes possible. 
The direction of motion of the charged cluster right after the tunneling, due to the Coulomb 
forces, will be away from the contact which has supplied the extra electron, 
and thus, we may say that the current becomes mechanically assisted. 
Hence, a sharp transition in the conductance of the system as 
the voltage increases beyond $V_{\mathrm{th}}^{01}$ is to be expected.

The same argument can be applied to the reverse jump, from the left island to the `$S$' electrode, 
\[
\mu_1^\leftarrow \simeq \frac{e^2(1-2x/d)}{2C_L(1+\delta_L\delta_R)} \left[ 
1 - \frac{V_S}{|V_{\mathrm{th}}^{01}|(1-x/d)} 
\right] . 
\] 
A negative chemical potential for the static island $\mu_1^\leftarrow<0$ requires now 
$V_S>0$ and $|V_S| > |V_{\mathrm{th}}^{01}|$.
If the voltage is just below the threshold, $V_{\mathrm{th}}^{01}(1-\Delta_\epsilon)$, a negative 
chemical potential involves as well $x/d>\Delta_\epsilon$, a process then exponentially suppressed.
 
It is easy to see that a similar situation occurs with the transport involving the right island and 
the right contact, since we have, by examining Eq. (\ref{eq:mus2}): 
\[
\mu_{i}^\rightleftarrows \simeq \frac{e^2(1-2x/d)}{2C_{\alpha_i}(1+\delta_L\delta_R)} \left[ 
1 \pm \frac{V_S}{V_{\mathrm{th}}^{0i}(1-x/d)} 
\right] , 
\]
with $i$ = 1, 3, $\alpha_1=L$, $\alpha_3=R$ and $V_{\mathrm{th}}^{03} = |e|\delta_R/2C_1$. 
Again, at zero temperature, a charge transfer from the right island to the right contact would 
require the island to separate from the contact when the voltage is just below the threshold 
$V_{\mathrm{th}}^{03}$. 
The $x$ dependency for $\mu_2^\rightleftarrows$ is not as important as in the 
other four processes, but naturally, the tunneling is favored as the islands 
approach each other, $x>0$. 

\begin{figure}[!hbt]
\centering\includegraphics[angle=0, width = 0.450\textwidth]{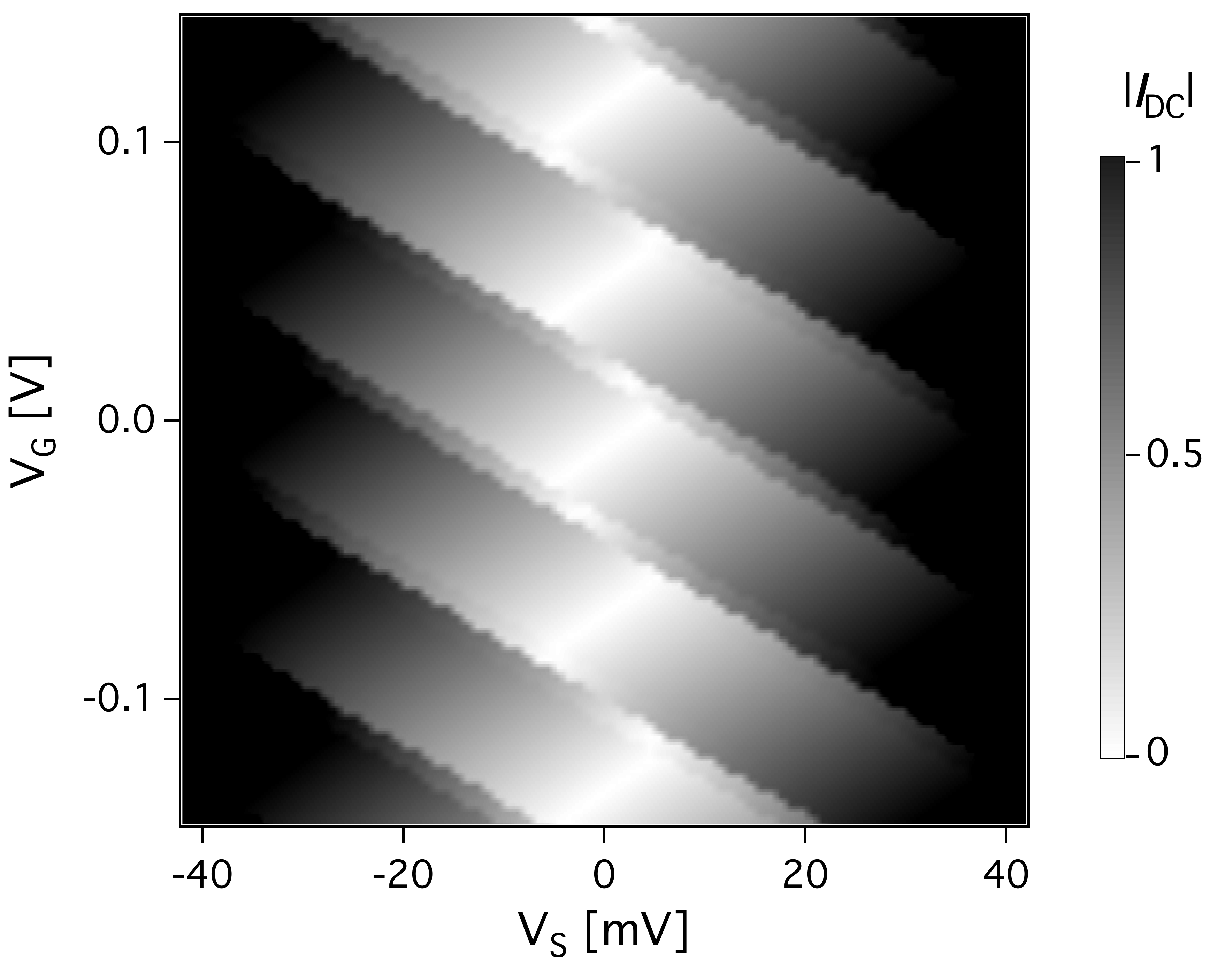}
\caption{\it \footnotesize
Contour plot of the absolute value of the direct current $|I_{\rm dc}|$ as a function of $V_S$ and $V_G$. 
The radii of the nanoislands were set to $r_L=28$ nm and $r_R = 31$ nm. 
Coulomb blockade diamonds are apparent, even at room temperature. 
}
\label{figIdc}
\end{figure}

We compute the current in the stationary limit, {\it i.e.}, when the transient 
solutions become negligible in the oscillations, 
\begin{equation}
I_{\mathrm{DC}} \sim -e 
\sum_{n_\alpha,n_\beta}P_{n_\alpha,n_\beta} \left [ 
\overrightarrow\Gamma^\alpha_{n_\alpha,n_\beta} - \overleftarrow\Gamma^\alpha_{n_\alpha,n_\beta}
\right ],
\label{eq:Idirect}
\end{equation}
where the non-equilibrium probability distribution $P_{n_\alpha,n_\beta}$ is a stationary solution
of the kinetic equation, (\ref{eq:master1}).
To perform our calculations, we used materials parameters mostly based on typical experimental values; see Table \ref{table1}. 
\begin{table}[!hbt]
\caption{\textit{Materials parameters used in this work. (PC = personal communication with Prof. Dr. Robert H. Blick, 
AE = authors estimation).
}}
\label{table1}\centering
\begin{tabular}{cccc}
\hline\hline
\ \ Parameter\ \  & \ \ \ \ Value \ \ \ \  &\  \ \ \ Units\ \ \ \ & \ \ Ref. \#  \ \ \\
\hline\hline
$\beta_1$ & 6.25  & $\mu$m$^{-1}$    & \cite{andrewsbook}   \\
\hline
$E$ & 150 & GPa & \cite{hopcroft} \& AE \\
\hline
$\gamma$ & .05  & \--\--   & PC  \\
\hline 
$R^0_{2}$ &  40 & $G\Omega$   & \cite{ck2}  \\
\hline
$R^0_{1,3}$ &  20 & $G\Omega$   & \cite{ck2}  \\
\hline
$C^0_{2}$ &  2 & aF &  \cite{ck2} \\ 
\hline
$C^0_{1,3}$ &  4 & aF &  \cite{ck2} \\ 
\hline 
$h$ & 250  & nm   & \cite{ck}, \cite{ck2}  \\
\hline
$r_R$ &  32 & nm &  \cite{ck2} \\ 
\hline 
$r_L$ &  28 & nm &  \cite{ck2} \\ 
\hline 
$d$ &  45 & nm & \cite{ck2} \\ 
\hline
$r_0$ &  7.5 & nm & AE \\ 
\hline
$L$ &  150 & nm & \cite{ck2} \\ 
\hline
$\lambda$ &  5 & nm & AE \\ 
\hline 
$m$ & 2$\times$10$^{-18}$ & Kg & \cite{ck}, \cite{ck2} \\
\hline
$\omega_0$ &  250 & MHz & \cite{ck} \\ 
\hline
$\epsilon$ & 10$^{-2}$ & \--\-- & AE  \\
\hline\hline 
\end{tabular}%
\end{table}
Figure \ref{figIdc} shows the numerical results of the absolute value of the normalized current $|I_{\rm dc}|$ 
using Eqs. (\ref{eq:master1}) and (\ref{eq:Idirect}), 
in units of $I_{\rm th}$, the current at the threshold bias, $V_{\rm th}$. 
The Coulomb blockade diamonds are apparent, a result of the discrete nature of the 
electronic charge \cite{vanderwiel,ck2}.

\begin{figure}[!hbt]
\centering\includegraphics[angle=0, width = 0.300\textwidth]{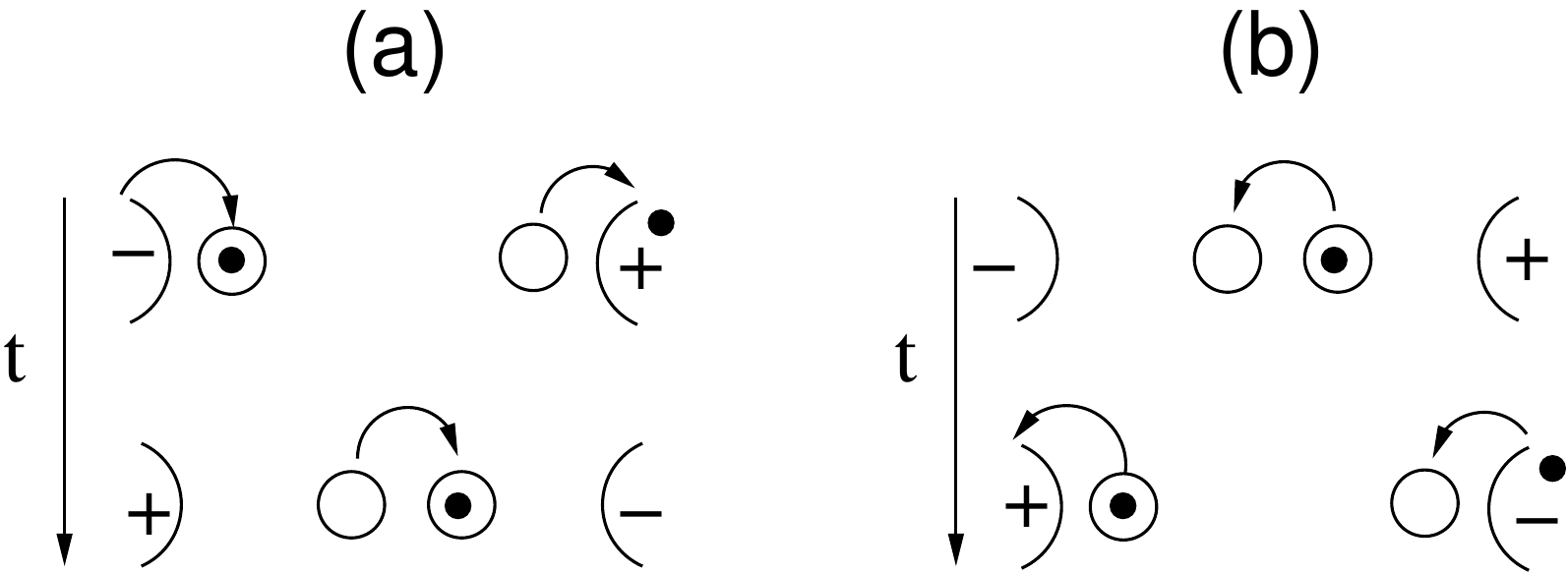}
\caption{\it \footnotesize
Electronic transport through a double pendula structure. 
The direct current is to the right (a) [left, (b)] direction if the metallic grains 
approach the contacts (each other) when the applied bias is negative. 
Electron transfer between the islands
to the right (left) direction occurs in (out of) phase with the signal, as depicted 
in bottom of (a) [top of (b)]. 
}
\label{figet}
\end{figure}

We now consider a rf signal superimposed  to the dc one, $V_S = V(t)$.  
The electron transfer between the islands 
to the right direction will occur {\em in phase} with the signal, and the 
electron transfer to the right direction between a contact and its nearest island
will occur {\em out of phase} with the signal (see Fig. \ref{figet}). 
The sequence of electronic transport is schematically depicted in Fig. \ref{figet}: 
The sign of the current is determined now by the initial conditions, occurring 
left to right in Fig. \ref{figet}(a) and right to left in Fig. \ref{figet}(b). 

If the frequency of the rf signal is close to the natural frequency of 
resonance $\omega_0$,  the oscillatory 
motion can assist the conductance by significant displacements of the island. 
The oscillatory regime in the frame of nonlinear dynamics is studied 
in the subsequent sections.  
\section{Electro-mechanical dynamics: Coupled Mode and Parametric Electronics}
\label{coupled}
As we noted before, the capacitances and the resistances depend on the 
displacements of the islands, hence affecting the chemical potentials and 
the tunneling processes, respectively. 
We want to take into account explicitly the oscillatory movement of the islands 
into the electrostatics.
To do so, we first express the free energy in terms of $E_{Ci}$ defined in 
Eq. (\ref{Eq:Eci}): 
\begin{eqnarray}
F(x;\{m_i\})&=& E_{CL}(x)\eta_L(\{m_i\}) + E_{CR}(x)\eta_R(\{m_i\})  \nonumber \\
&+&E_{CC}(x)\eta_C(\{m_i\}),
\end{eqnarray}
where we have defined the set of variables $\eta$ as: 
\begin{eqnarray}
\eta_L &=& \frac{n_L}{2} - \frac{m_1}{|e|}\left[V_GC_{G1} + V_S(C_L-C_1) \right] 
-\frac{m_2}{|e|}\left[V_SC_1 \right]
\nonumber \\
\eta_R &=& \frac{n_R}{2} 
+\frac{m_3}{|e|}\left(V_GC_{G2} \right)
\nonumber \\
\eta_C &=& n_Ln_R - \frac{m_1}{|e|}\left[V_GC_{G2} + V_SC_2^0 \right] 
+\frac{m_2}{|e|}\left(V_SC_1 \right) +\nonumber \\ &+&
\frac{m_3}{|e|}\left(V_GC_{G1}+V_SC_1 \right)
\nonumber 
\end{eqnarray}
Next, we expand $E_{Ci}$ in terms of the relative displacement $x_d = x/d$ (note that $x_d<1$): 
\begin{eqnarray}
E_{CL/R}&\simeq &\frac{e^2}{8\epsilon r_{L/R} (1-\delta_r)}
\left[ 1 +
\frac{2\delta_r}{1-\delta_r} x_d +  
\frac{3\delta_r+\delta_r^2}{ (1-\delta_r)^2} x_d^2 +
\dots
\right] \nonumber\\
E_{CC}&\simeq &\frac{e^2}{8\epsilon d (1-\delta_r)}
\left[ 1- \frac{1+\delta_r}{1 - \delta_r} x_d 
+ \frac{1 +3\delta_r}{(1-\delta_r)^2} x_d^2
+\dots
\right],\nonumber\\ 
\label{eq:devU}
\end{eqnarray}
where $\delta_r = r_Lr_R/d^2\ll 1$. 
We include the electrostatic force in the coupled harmonic oscillators
by adding the term $F(n_L,n_R;x)$ to the Lagrangian, 
$\mathcal L = m\dot{x}^2/2 -kx^2/2 - F(n_1,n_2;x)$. 
Using the leading terms in $\delta_r$ for 
the derivative of Eq. (\ref{eq:devU}), we have: 
\begin{eqnarray}
-\frac{\partial F}{\partial x} &\simeq &
-\frac{e^2}{ 8\epsilon d} 
\left\{ \left(\frac{\eta_L}{r_L} + \frac{\eta_R}{r_R} \right)\delta_r
\sum_{n=1}^\infty \frac{n(n+1)x_d^{n-1}}{(1-\delta_r)^n } + 
\right.\nonumber \\ && \left.
 + \frac{\eta_C}{d} \sum_{n=1}^\infty \frac{nx_d^{n-1}}{(1-\delta_r)^n}
\right\} 
\end{eqnarray}
The Lagrange's equation of motion for the relative coordinate $x$ now reads: 
\begin{equation}
m \ddot x +m\gamma\dot{x} + k x 
= -\frac{\partial F}{\partial x} \to 
m \ddot x +m\gamma\dot{x} +(k+\Delta_k) x = -\Delta_F
\label{eq:mathieu}
\end{equation}
where we have introduced the damping force,  $F_\gamma = m\gamma \dot x$ and defined
\begin{eqnarray}
\Delta_k &=& \frac{e^2}{4\epsilon d^2(1-\delta_r)^2}\left[ 3\delta_r 
\left(\frac{\eta_L}{r_L} + \frac{\eta_R}{r_R} \right) 
+ \frac{\eta_C}{d} \right]; \quad 
\nonumber \\
\Delta_F &=& - \frac{e^2}{8\epsilon d(1-\delta_r) }\left[ 
2\delta_r \left(\frac{\eta_L}{r_L} + \frac{\eta_R}{r_R} \right)
+  \frac{\eta_C}{d} 
\right]. \nonumber
\end{eqnarray}
Equation (\ref{eq:mathieu}) is known as a damped Mathieu equation \cite{mathieu}. 
It includes a standard harmonic oscillator driving term, and a {\it parametric modulation} term,
which is a variable spring constant, $\Delta_k$.
We note that $\Delta_k$ and $\Delta_F$ depend in general on time, as the shuttle oscillates
causing a charge transfer. 
A rough estimation for a realistic system 
gives $\Delta_F\simeq0.2-1$ pN.
$\eta_{L,R,C}$ depend on the number of excess electrons on each of the islands, $n_{L/R}$. 
As we will see below, in the shuttling regime, $n_{L/R}$ is a function that oscillates with 
time, as the mechanical movement of the islands assist the electronic transport through 
the device. Thus, $\Delta_k$ and $\Delta_F$ are time-dependent functions, and Eq. (\ref{eq:mathieu})
is a modified Mathieu equation, which can be treated numerically \cite{pratero}. 
In the following subsections, we consider two different limits to understand the
dynamics of the system. 
\subsection{Small oscillations limit in the linear regime}
We consider first the small oscillations within the classical circuit limit. 
In the linear adiabatic regime, the charge balance in the islands follows the excitation given by 
the applied voltage, $V(t)$. 
From classical circuit theory, we have that 
the applied voltage equals to the sum of the voltages that drop on each junction, and
the net current through each of the junctions is the same,
\begin{equation}
V_{SD}= V(t) = \sum_{i=1}^3\frac{q_i}{C_i}; \quad 
\frac{q_i}{R_iC_i} = \frac{q_j}{R_jC_j}. 
\label{eqconds}
\end{equation}
We can solve the above system of equations for $q_i = -|e|m_i$ and get
the charge on each island, $Q_i = -|e|n_i$: 
$Q_L=q_1-q_2$; $Q_R=q_2=q_3$.
When the flexural modes of the nanopillars are excited, the mutual
capacitance and the resistances become sensitive to the displacements,
$x_L$ and $x_R$. In a flexural mode in which the center of mass is at rest ($X_0=0$), as in 
Fig. \ref{fig1abc}(d), 
the resistances are given by (\ref{eq:resis}). The mutual capacitance $C_2$ determining 
the coupling of the metallic islands depends as well in their separation,
whereas $C_{1,3}$ can be considered as constant, 
\[
C_{1,3} \simeq C_{1,3}^0; \quad C_2(x) = \frac{C_2^0}{1-x/d}.\nonumber 
\]
Further, we will make the approximation that $R_i^0C_i^0$ is a constant, but the
resistance $R_2$ is twice the resistances $R_{1,3}$. We set
$R_1^0 = R_3^0 = R_2^0/2$ and $C_1^0=C_3^0=2C_2^0 = C$, consistent 
with previous results \cite{ck2}.
From now on, we express the relative coordinate in units of $\lambda$,  
$x\equiv x/\lambda$. Using (\ref{eqconds}) 
we get 
\[
q_2 = \frac{CV(t)}{2(1-x/d)(1+e^{3x/2} ) }; \quad
q_1=-q_3 = \frac{CV(t)e^{3x/2}}{2(1+e^{3x/2} ) }, 
\]
giving a compact expression for $Q_{L} = -Q_R$,
\begin{equation}
Q_L\simeq \frac{CV}{2}
\left[
\tanh{\frac{3x}{4}}-\frac{x}{d} e^{{3x/4}}
\right]
\simeq 
\frac{3CV}{4}\left[
x- \frac{2}{d}x^2-\frac{3}{16}x^3
\right]. 
\label{eqQ}
\end{equation}
At lowest order in $x$, we have a symmetric system
whose islands have no net charge at $x=0$. Then, for $x<0$ (so the
pillars are away from each other, getting close to the contacts), a net charge
with negative sign is induced in island L and a positive one in R. Note
that the potential is, by convention, negative on the left and positive on
the right contact. Reversing the potential $V$ will cause a change of sign in
$Q_L$ and $Q_R$, as expected.
The two nonlinear terms on the right break the left-right symmetry of the system. 

The dynamics of the nanopillars will consist of a set of oscillators experiencing
the electric field,  $V(t)/L$. 
We follow the approach given by Ahn {\it{et al.}} \cite{ahn} to investigate
the electrodynamics of the system. 
We find the equations of motion of the relative coordinate 
$x$ by setting $m=m_R\simeq m_L$ and substituting Eq. 
(\ref{eqQ}) 
into Eq. (\ref{eq:mathieu}),
\[
\ddot x+ \frac{\gamma}{\omega_0} \dot x+ x= -\alpha \sin^2\omega \tau
\left[ {x} - \frac{2\lambda }{d}x^2 - \frac{3}{16}x^3 \right].
\]
Here 
$\alpha = 3CV_0^2/4Lm\lambda\omega_0^2$, a dimensionless forcing parameter 
which account for the ratio of the electric ($F_e\sim CV_0^2/L$) and mechanical forces
($F_m\sim m\lambda\omega_0^2 = k\lambda$). 
We note that for a typical Si nanopillar, $F_m \sim 50-60$ pN and $F_e\sim 1-5$pN. 
We have also rescaled the time, $\tau = \omega_0 t$. 
$\omega$ is the frequency of the excitation, expressed in units of $\omega_0$. 
This is a non-linear equation that corresponds to a forced and damped oscillator,
where the forcing terms depend on the coordinate itself. 
At first order in $x$, we obtain a modified Mathieu equation, which
gives instability regions when the excitation is strong enough.
We are, however, interested also in the weak excitation regime, in which the non-linear
terms and non-linear effects such as Coulomb blockade could play a critical role.

Following  the Poincar\'e-Lindstedt method, 
we parametrize the damping and the forcing using a small arbitrary $\epsilon$ ($\epsilon \ll 1$), 
$\gamma\sim \epsilon \gamma_1$, $\alpha \sim \epsilon\alpha_1$, 
\begin{equation}
\ddot x + x + \epsilon\left( \gamma_1{\dot x} +   \alpha_1\sin^2{\omega \tau} 
\left[ x- \frac{2\lambda }{d}x^2  -\frac{3}{16}x^3 \right]
\right) = 0.
\label{eqnldyn}\end{equation}
Thus, we can consider two different time scales,
the ``stretched'' time, $z = \omega \tau$, 
and the ``slow'' time, $\eta = \epsilon \tau$. 
The time derivatives are now expressed in terms of these new times as
\begin{equation}
\dot x = 
\omega\frac{\partial x}{\partial z} + \epsilon \frac{\partial x}{\partial \eta}; 
\quad
\ddot x = 
\omega^2\frac{\partial^2 x}{\partial z^2} 
+ 2\epsilon 
\omega\frac{\partial^2 x}{\partial \eta \partial z} 
+ \epsilon^2\frac{\partial^2 x}{\partial \eta^2} . 
\label{eqexx}
\end{equation}
We expand $x$ in terms of $\epsilon$, 
\begin{equation}
x(z,\eta) \simeq x_0 + \epsilon x_1 + \dots \quad 
\label{eqexw}
\end{equation}
and, likewise, seek for solutions that correspond to harmonics of the
natural frequency $\omega_0$:
\begin{equation}
\omega \simeq p + \epsilon \delta_\omega + \dots, 
\label{eqew}
\end{equation}
where $p$ is an integer or fractional number
and, finally, substitute (\ref{eqexx}), (\ref{eqexw}), and (\ref{eqew}) into (\ref{eqnldyn}), neglecting terms
of $O(\epsilon^2)$, which gives, after collecting terms at lowest order in $\epsilon$,
\[
\frac{1}{p^2}\frac{\partial^2x_0}{\partial z^2} + x_0 = 0  
\]
giving a general solution for $x_0$, 
\begin{equation}
x_0(z,\eta) = A(\eta)\cos{\frac{z}{p}} + B(\eta)\sin{\frac{z}{p}} .
\label{eqsol}
\end{equation}
The constants of integration, $A$ and $B$ are functions of the ``slow'' time $\eta$.
At first order in $\epsilon$ we get:
\begin{eqnarray}
\frac{\partial^2x_1}{\partial z^2} + x_1  
&=&-2\frac{\partial^2x_0}{\partial z\partial \eta} 
-2\delta_\omega\frac{\partial^2x_0}{\partial z^2} -\gamma_1 \frac{\partial x_0}{\partial z} - 
\nonumber \\ &-&
\alpha_1 (1-\cos{2z} )
\left[
x_0-\frac{2\lambda x_0^2}{d}-\frac{3x_0^3}{16}  \right]. 
\label{eqx1}
\end{eqnarray}
Without loss of generality, we may ask that $x_1$ satisfies $\ddot x_1 + x_1 = 0$. 
We substitute the general solution (\ref{eqsol}) for $p =1$ into the above
expression and arrange terms in $\sin{z}$ and $\cos{z}$ (see Appendix \ref{app1}) to get:
\begin{eqnarray}
2\frac{\ud A }{\ud \eta} &= &
-\gamma_1 A -\left(2\delta_\omega - \frac{3\alpha_1}{2}\right)B- 
\frac{3\alpha_1}{64}B(3A^2 + 5B^2); \nonumber \\
2\frac{\ud B }{\ud \eta} &= &
-\gamma_1 B +\left(2\delta_\omega - \frac{\alpha_1}{2}\right)A+ 
\frac{3\alpha_1}{64}A(A^2 + 3B^2).
\nonumber \\
\label{eqAB}
\end{eqnarray}
Note that equilibrium points of (\ref{eqAB}) correspond to periodic solutions of our forced
oscillator, and 
the norm of the solutions is conserved, {\it i.e.}: $A^2+B^2 = x_0^2+\dot x_0^2$.

If a dc signal is superimposed, 
$V(t) = V_0(\sin{\omega \tau} + \beta)$ with  $\beta = V_{dc}/V_{0}$,
Eq. (\ref{eqnldyn}) will now read
\begin{eqnarray}
&&\epsilon\left\{ \gamma_1\dot x + 
\alpha_1[\sin^2\omega \tau+2\beta\sin\omega \tau+\beta^2]
\left[ {x} - \frac{2\lambda }{d}x^2  -\frac{3}{16}x^3 \right]
\right\}  \nonumber \\&& 
+\ddot x + x  = 0.
\nonumber
\label{eqnldyn3}
\end{eqnarray}
Proceeding in the same manner as before, we find six extra terms (terms in $\beta$ or $\beta^2$).
We consider the stretched and slow time, and
expand $x$ and $\omega$ in terms of $\epsilon$, to get, for the $p$ = 1 case (see Appendix \ref{app2}):
\begin{widetext}
\begin{eqnarray}
\frac{\ud A }{\ud \eta} &= &
-\gamma_1 A -\left(\delta_\omega - \frac{3\alpha_1}{4} - {\alpha_1 \beta^2} \right)B- 
\frac{\alpha_1\beta\lambda}{d}(A^2+3B^2)-\frac{27\alpha_1\beta^2}{64}B(A^2+B^2)-
\frac{3\alpha_1}{128}B(3A^2 + 5B^2)\nonumber \\
\frac{\ud B }{\ud \eta} &= &
-\gamma_1 B +\left(\delta_\omega - \frac{\alpha_1}{4}- {\alpha_1 \beta^2}\right)A
+\frac{2\alpha_1\beta\lambda}{d}AB+\frac{27\alpha_1\beta^2}{64}A(A^2+B^2)
+\frac{3\alpha_1}{128}A(A^2 + 3B^2)
\nonumber\\
\label{eqABDC}
\end{eqnarray}
\end{widetext}
and for the $p$ = {\emph ``any''} case, (see Appendix \ref{app1b}):
\begin{widetext}
\begin{eqnarray}
2\frac{\ud A }{\ud \eta} &= &
-A\left(\gamma_1-\beta\alpha_1\delta_{p,2}\right) - 
B \left(\frac{2\delta_\omega}{p} - \frac{\alpha_1}{4}(2+  \delta_{p,1}+4\beta^2)\right)+ 
AB\frac{\alpha_1\lambda}{8d}\delta_{p,3/2} + (A^2-B^2)\frac{\beta\alpha_1\lambda}{4d}\delta_{p,3}
\nonumber \\ && 
-\frac{3\alpha_1}{256}\left[ B(A^2+B^2)(6+12\beta^2 +3\delta_{p,1}) + 
   B(3A^2-B^2)\delta_{p,2} +6A(A^2+B^2)\delta_{p,2} + 2A(A^2-3B^2)\delta_{p,4}  \right] 
\nonumber \\
2 \frac{\ud B }{\ud \eta} &= &
-B\left(\gamma_1+\beta\alpha_1\delta_{p,2}\right) + 
A \left(\frac{2\delta_\omega}{p} - \frac{\alpha_1}{4}(2-  \delta_{p,1}+4\beta^2)\right)+ 
(A^2-B^2)\frac{\alpha_1\lambda}{16d}\delta_{p,3/2}  
- AB\frac{\beta\alpha_1\lambda}{2d}\delta_{p,3}
\nonumber \\ && 
+\frac{3\alpha_1}{256}\left[ A(A^2+B^2)(6+12\beta^2 -3\delta_{p,1}) - 
   A(A^2-3B^2)\delta_{p,2} +6B(A^2+B^2)\delta_{p,2} + 2B(3A^2-B^2)\delta_{p,4}  \right] 
\nonumber\\
\label{eqAB2}
\end{eqnarray}
\end{widetext}
The stability of the solutions of the equation above can be investigated using 
numerical methods \cite{pratero}.

\begin{figure}[!hbt]
\centering\includegraphics[angle=0, width = 0.450\textwidth]{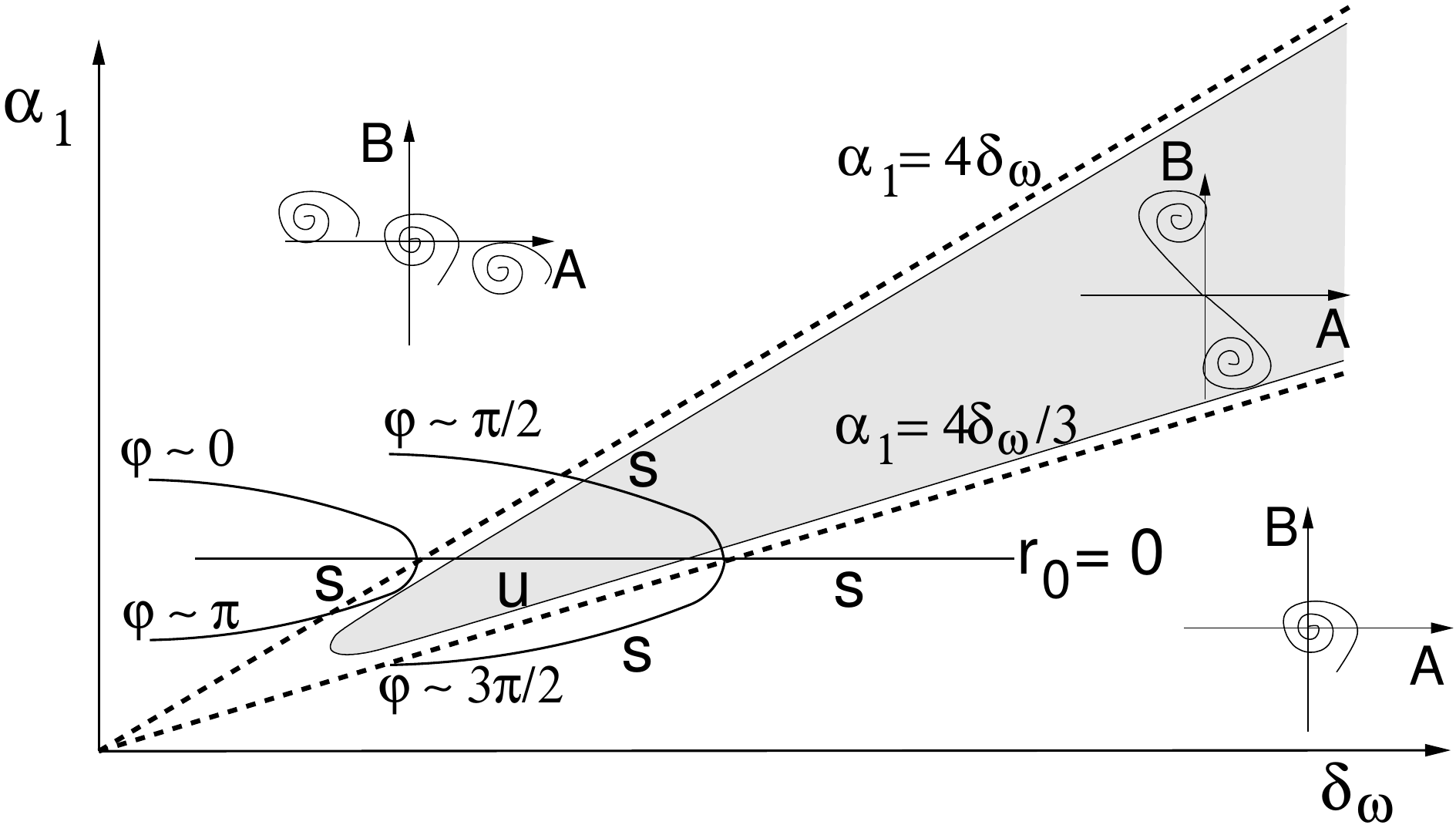}
\caption{\it \footnotesize 
Schematic representation of the 
$p$ = 1 tongue in the parameter space spanned by $\delta_\omega$ and $\alpha_1$. 
Between the transition curves (broken lines), the trivial solution is unstable and two 
stable solutions are born, $\varphi$ = 0, $\pi$ (solid curves). 
The insets depict phase portraits in the $A$-$B$ plane: 
below the lower transition curve, only the trivial solution exists. In between both 
transition curves, the trivial solution is unstable, and another two solutions are born. 
Above the second transition curve, $r_0$ = 0 is stable again, and the other two solutions
become unstable.
} 
\label{fig:paraplane}
\end{figure}

In order to understand qualitatively the electromechanical motion scenario, 
we consider, first, the case $p$ = 1 and $\beta$ = 0, where Eq. (\ref{eqABDC}) reads
($\dot A \equiv \ud A/ \ud \eta$)
\begin{eqnarray}
2\dot A  & = & -\gamma_1 A -(\gamma_2-\gamma_3)B - \gamma_4 B(3A^2 + 5B^2)\nonumber \\  
2\dot B  & = & -\gamma_1 B +(\gamma_2+\gamma_3)A + \gamma_4 A(A^2 + 3B^2)\nonumber 
\end{eqnarray}
where we have defined 
$\gamma_2 = 2\delta_\omega -\alpha_1$, 
$\gamma_3 = \alpha_1/2$, and 
$\gamma_4 = 3\alpha_1/64$. 
It is easy to see that the transition curves for the stability of the trivial 
solution are $\gamma_2 = \pm \gamma_3$, or 
$\alpha_1 = 4\delta_\omega$,  $4\delta_\omega/3$. Along this curves (broken lines 
of Fig. \ref{fig:paraplane}) the stability of $r_0$ = 0 changes. 
To gain in simplicity, we transform to polar coordinates in the $A$-$B$ phase plane, 
by setting $A = r_0 \cos{\varphi}$ and  $B = r_0\sin{\varphi}$. The amplitude 
$r_0^2 = A^2 + B^2$ and the phase $\varphi = \arctan{B/A}$ now satisfies: 
\begin{eqnarray}
2\dot r_0 &= &-\gamma_1 r_0  + r_0 \sin{2\varphi}(\gamma_3  - \gamma_4 r_0^2) \nonumber \\
2\dot \varphi &=& \gamma_2 +\gamma_3 \cos{2\varphi} + \gamma_4 r_0^2 (3-2\cos{2\varphi})
\label{eq:polar}
\end{eqnarray}
We seek equilibria of the ``slow flow'' (\ref{eq:polar}). A solution in which $r_0$ and $\varphi$ 
are constant represents a periodic motion of the nonlinear Mathieu equation, which has the
frequency of the forcing function. Such equilibria satisfy $\dot r_0$ = $\dot\varphi$ = 0.  
Ignoring the trivial solution $r_0$ = 0, the first equation of (\ref{eq:polar}) with 
$\dot r_0$ = 0 requires 
\[
\sin{2\varphi} = \frac{\gamma_1}{\gamma_3 - \gamma_4 r_0^2}. 
\]
In the absence of damping $\gamma_1\sim 0$, find equilibria at $\varphi$ = 0, $\pi/2$, $\pi$, $3\pi/2$. 
The second equation of (\ref{eq:polar}) with $\dot \varphi$ = 0 then implies 
\[
r_0^2 = -\frac{\gamma_2 + \gamma_3 \cos{2\varphi}}{\gamma_4 (3-2\cos{2\varphi}} = -\frac{\gamma_2 \mp \gamma_3}{\gamma_4 (3\mp 2)}. 
\]
For a nontrivial real solution, $r_0^2>0$. In the case of $\varphi$ = 0 or $\pi$, $\cos{2\varphi}$ = 1
and nontrivial equilibria require $-\gamma_3 - \gamma_2>$ 0 or $4\delta_\omega<  \alpha_1 $.
On the other hand, for $\varphi$ = $\pi/2$ or $3\pi/2$, $\cos{2\varphi} = -1$
and nontrivial equilibria require $\gamma_3 - \gamma_2>$ 0 or $4\delta_\omega<  3\alpha_1 $. 
Since $\delta_\omega = \alpha_1/4$ and $\delta_\omega = 3\alpha_1/4$ correspond to 
transition curves for the stability of the trivial solution, bifurcations occur as 
we cross the transition curves in the $\delta_\omega$-$\alpha_1$ plane 
(see Fig. \ref{fig:paraplane}). 
Keeping $\alpha_1$ fixed as we shift $\delta_\omega$ from the right across the right 
transition curve, the trivial solution $r_0$ = 0 becomes unstable and simultaneously two branches of stable solutions  
are born, one with $\varphi \sim \pi/2$ and the other with $\varphi \sim 3\pi/2$. 
This motion grows in amplitude as $\delta_\omega$ continues to decrease. 
When the left transition curve is crossed, the trivial solution becomes stable again. 
This scenario can be pictured as involving two pitchfork bifurcations, as depicted in 
Fig. \ref{fig:paraplane}.  

A finite damping $\gamma_1$ shifts slightly the position of 
the stable equilibria in the $A$-$B$ plane, which become stable spirals, as 
represented in the insets of Fig. \ref{fig:paraplane}. The damping also 
``shrinks'' the region of instability (shaded area of Fig. \ref{fig:paraplane}), 
lifting it away from the origin in the parameter space. 

In the classical limit, we can have an idea of the resulting direct 
current through the system. The time-average direct current is obtained 
by integrating over a period the current across one of the three junctions of 
Fig. \ref{fig1},
\begin{equation}
I_{\rm dc} = \frac{\omega}{4\pi R}\int_{t_0}^{t_0+T}\frac{V(t)e^{x}}{1+e^{3x/2}} 
.
\label{eq:Idc}
\end{equation}
We study (\ref{eq:Idc}) in terms of the coefficients $r_0$ and $\varphi$, 
$x  = r_0 \cos{(\omega t - \varphi)}$. We
find that the absolute value of the direct current reaches a 
maximum when $\varphi = 0, \pi$ and $r_0 = 2$.  

\begin{figure}[!hbt]
\centering\includegraphics[angle=0, width = 0.450\textwidth]{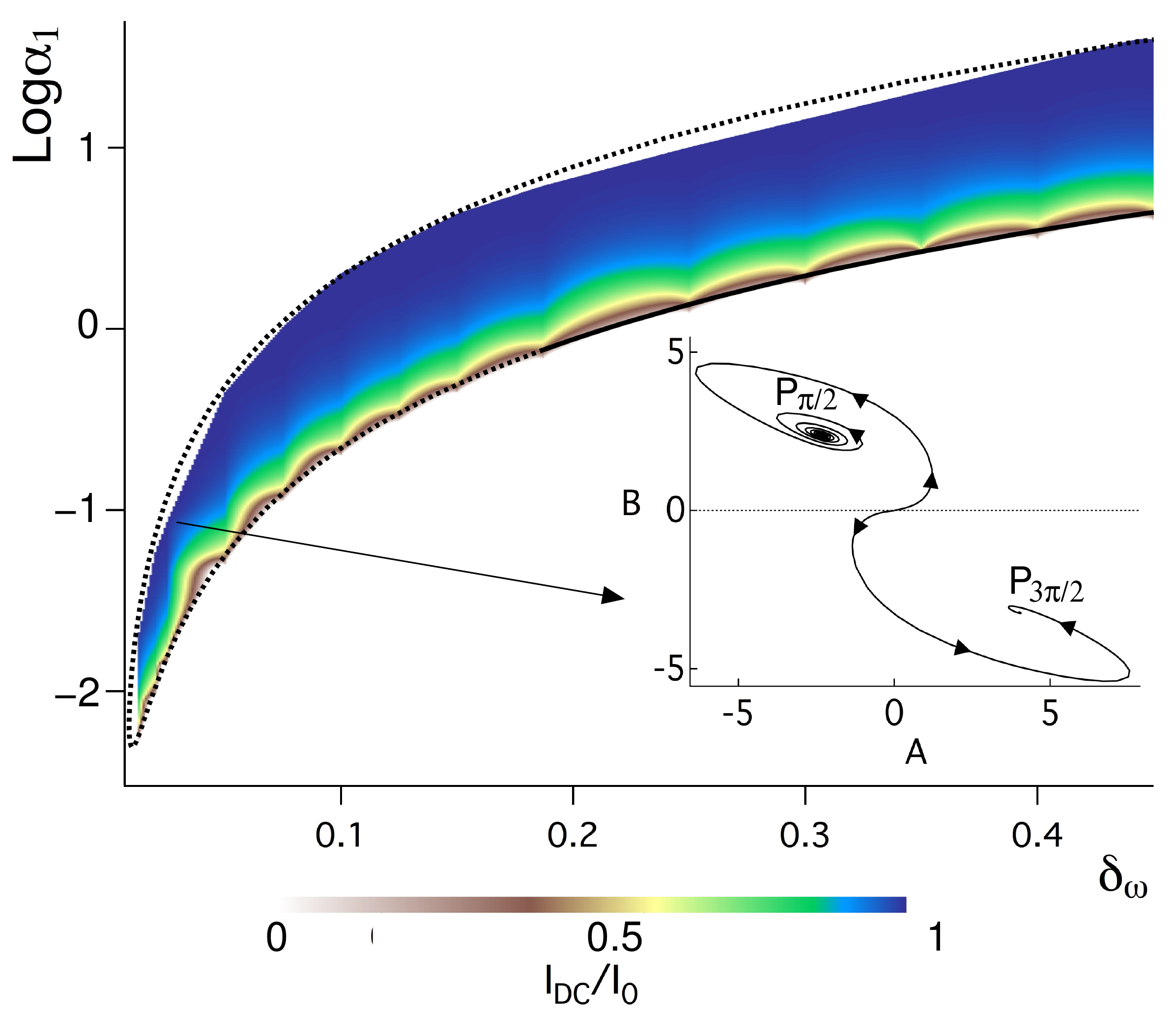}
\caption{\it \footnotesize (Color online) Contour plot of the direct current 
$I_{\rm dc}$ as a function of $\alpha_1$ and $\delta_\omega$ in the first tongue, $p =1$. 
The inset shows a phase portrait in the $A$-$B$ plane: inside the tongue, 
the origin is unstable, and two stable solutions are 
found, $P_{\pi/2}$ and $P_{3\pi/2}$.  
}
\label{fig:IandAB}
\end{figure}
Figure \ref{fig:IandAB} shows numerical results of $I_{\mathrm dc}$ in the 
$p = 1$ tongue, where $\alpha_1$ is in log scale. Inside the tongue, the 
trivial solution is unstable, and two stable solutions appear at 
$P_{\pi/2}$ and $P_{3\pi/2}$ in the $A$-$B$ plane (see inset). 
The trajectories stay either in the upper or lower semiplane; thus, 
the initial conditions determine the point of stability: If $\varphi(0)>0$ ($<0$), 
then the phase portrait in the $A$-$B$ plane reaches $P_{\pi/2}$ 
($P_{3\pi/2}$) and the electron transport is right to left (left to right), 
using the convention of Fig. \ref{figet}.  
At these points, the nanopillars are oscillating with the natural frequency 
of the oscillations, mechanically assisting the electronic transport.
\subsection{Oscillations in the shuttling regime}
So far we have studied the limit when the charge in the metallic islands 
is given by Eq. (\ref{eqQ}), {\it i.e.}, the charge on the islands 
$Q_{L,R}$ changes continuously, following the excitation $V(t)$. 
As the size of the metallic islands shrinks, however, we may reach the 
discrete limit, 
where single-electron effects such as Coulomb blockade become important.  
In the Coulomb blockade limit, as we have seen in Sec. \ref{transport},
an extra electron can only be added to the island if enough energy is provided 
by the external sources to overcome the Coulomb repulsion between the electrons. 
The equation of motion for the relative coordinate reads: 
\begin{equation}
\label{eq:dyn}
\ddot{x} + \gamma \dot x+ x= \frac{e V(t)}{kL\lambda} n(t),
\end{equation}
where $n(t) = n_L(t)-n_R(t)$. 

\begin{figure}[!hbt]
\centering\includegraphics[angle=0, width = 0.45\textwidth]{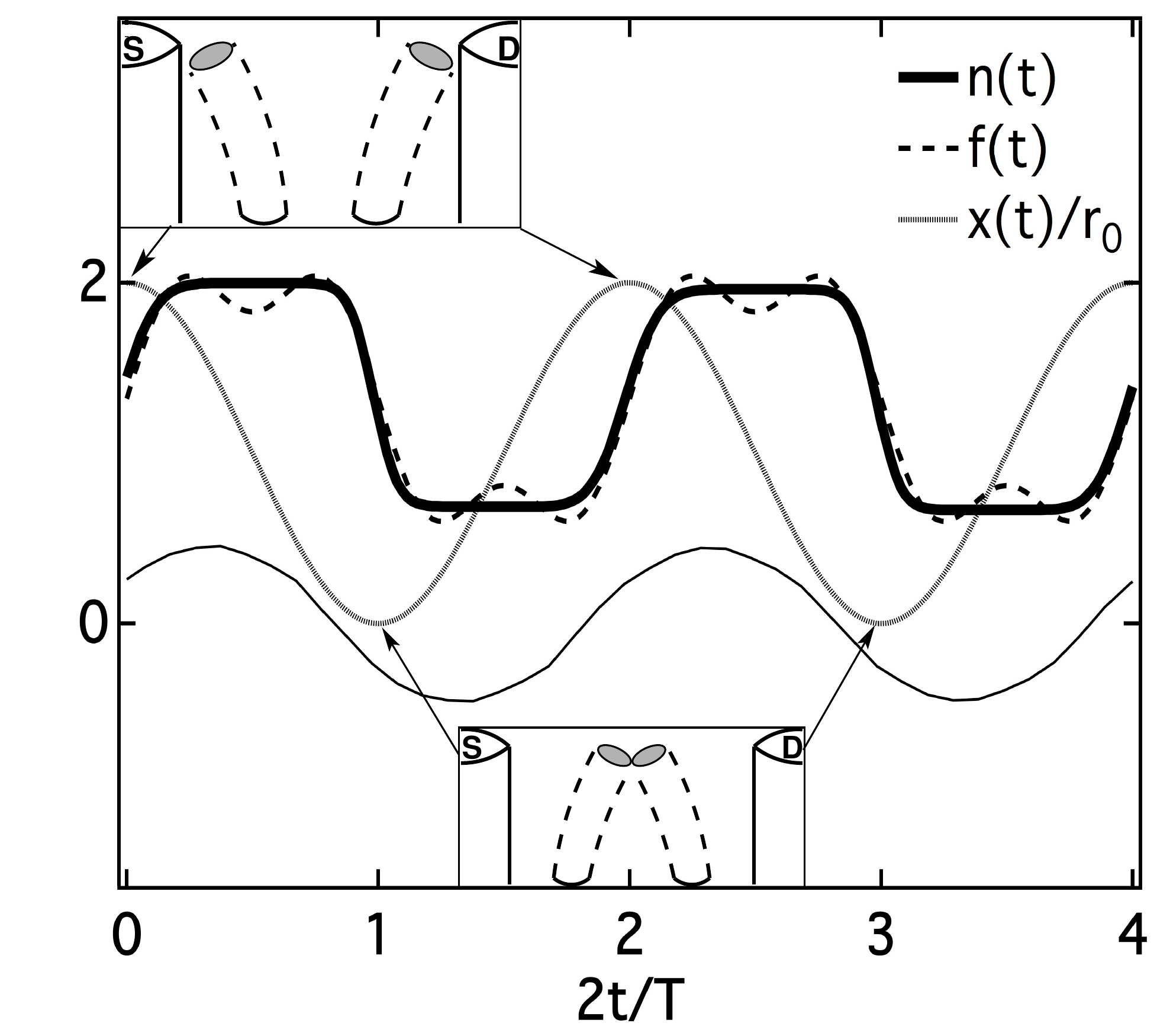}
\caption{\it \footnotesize 
Time evolution of $n(t) = n_l - n_R$ in the stationary limit for low (thin solid curve)
and high bias (thick solid curve). In the high bias regime, $n(t)$ can be 
approximated by a square wave. 
}
\label{fign}
\end{figure}
The evolution of the probabilities of having $n_L$ excess electrons on the left island and
$n_R$ on the right island is given by Eq. (\ref{eq:master1}) , which is solved by direct integration. 
Figure \ref{fign} shows numerical results of the function $n(t)$ obtained solving 
simultaneously the master equation [Eqs. (\ref{eqGamma})-(\ref{eq:master1})] and the dynamic equation, [Eq. \ref{eq:dyn}] 
in the stationary regime. 
In the low bias limit, $n(t)$ is a sinusoidal function that follows the excitation 
$V(t)$ (thin solid curve of Fig. \ref{fign}).  
The charge has only a small probability to be transferred across the device, 
following the bias. 
However, after some critical bias, the charge transfer occurs mostly at the points 
of maximal deflection, commonly termed as shuttling regime. 
We obtain numerically $n(t)$, resulting a square wave correlated with the excitation $V(t)$ 
(thick solid curve of Fig. \ref{fign}), 
\begin{equation} 
n(t) \simeq n_{\mathrm{av}} + 4n_0(\cos{\omega_0t} - \cos{3\omega_0t})/\pi, 
\label{eq:n}
\end{equation}
where $n_{\mathrm{av}}$ and $n_0$ are obtained after averaging over a large number of simulations 
and depend on the input parameters. 
We note that the charge transfer occurs in the points of maximal deflection during an 
effective contact time, in accordance with Weiss {\it et al.} \cite{weiss_zwerger}. 

Inserting the expression (\ref{eq:n}) into (\ref{eq:dyn}), 
we aim, as before, for oscillatory solutions,
$x_0 \simeq A(\eta) \cos{z/p} + B(\eta)\sin{z/p}$,
bearing in mind the following linearized equations for the coefficients $A$ and $B$ (see Appendix \ref{app:CB}):
\begin{eqnarray}
\label{eq:AB_CB}
2\frac{\ud A }{\ud \eta} &= &
-\gamma_1{A} -  \frac{2\delta_\omega}{p}  B
+ 
n_{\mathrm{av}}\alpha_1^\prime \delta_{p,1} + 
n_0\frac{\alpha^\prime_1}{6}(2\delta_{p,2} - \delta_{p,4})
\nonumber \\
2 \frac{\ud B }{\ud \eta} &= &
-\gamma_1{B} +  \frac{2\delta_\omega}{p} A
+ \alpha_1^\prime\beta n_0, 
\end{eqnarray}
where we have defined $\alpha^\prime = eV_0/L k \lambda$ and
$\alpha^\prime_1  = \epsilon \alpha^\prime$. 
As before, $\alpha^\prime$ can be viewed as the ratio between the electrical and mechanical forces. 
We find the equilibrium or stable points by solving  Eq. (\ref{eq:AB_CB}) with $\dot A$ = $\dot B$ = 0. 
In the absence of $V_{{\mathrm{dc}}}$ ($\beta = 0$), non-trivial stable points are found
only around subharmonics with $p$ = 1, 2, or 4, 
\begin{eqnarray}
\label{eq:sp1}
(A_{\mathrm{eq}}, B_{\mathrm{eq}})_{p=1} &=&  
\frac{\alpha^\prime n_{\mathrm{av}}}{\gamma_1^{2} +  4\delta_\omega^2} (\gamma_1,{2\delta_\omega})
\nonumber \\
 (A_{\mathrm{eq}}, B_{\mathrm{eq}})_{p=2} &=& 
\frac{\alpha^\prime n_{\mathrm{0}}}{3(\gamma_1^{2} +  \delta_\omega)^2} (\gamma_1,{\delta_\omega})
\nonumber \\
 (A_{\mathrm{eq}}, B_{\mathrm{eq}})_{p=4} &=& 
\frac{\alpha^\prime n_{\mathrm{0}}}{6(\gamma_1^{2} +  (\delta_\omega/2)^2)} \left(\gamma_1,\frac{\delta_\omega}{2}\right)
\end{eqnarray}
In contrast, under a finite dc bias, nontrivial stable points are found for {\em any} frequency
at 
\begin{equation}
\label{eq:sp}
(A_{\mathrm{eq}}, B_{\mathrm{eq}})_{p} =
\frac{ \alpha\beta n_0}{ \gamma_1^{2}+(2\delta_\omega/p)^2 } \left(\frac{2\delta_\omega}{p}, -\gamma_1\right). 
\end{equation}
We stress that this result is valid in the limit of large oscillations.
To estimate the range of validity of Eqs. (\ref{eq:sp1}) and (\ref{eq:sp}), 
we consider the limit of small oscillations, for which the charge on the islands 
``follows'' the mechanical motion, 
with an amplitude proportional to the amplitude of the oscillations, $r_0 = \sqrt{A^2+B^2}$, 
\[
n(t) \simeq a_n r_0 \sin{\omega_0 t}. 
\]
In this limit, Eq. (\ref{eq:AB_CB}), in polar coordinates and for $\beta\neq 0$ and $p\neq 2$ reads 
\[
\dot r^2_0 = -\frac{1}{\gamma_1} r_0^2 + \alpha_1^\prime \beta  a_n A r_0 , 
\]
We find a change in the stability of $r_0 = 0$ when the forcing parameter exceeds a threshold, 
$\alpha_1^\prime \beta a_n > \alpha_{\mathrm{th}}^\prime$, with 
\[
\alpha_{\mathrm{th}}^\prime = \sqrt{\gamma_1^{2} + \delta_\omega^2}.
\]
$r_0$ changes from a stable to an unstable spiral as 
the forcing is increased beyond $\alpha_{\mathrm{th}}^\prime/(\beta a_n)$. 
In this situation, the amplitude of the oscillations could reach the Fowler-Nordheim 
tunneling limit \cite{scheible_prl}, with a subsequent enhancement of the direct current 
due to field emission, marking the limit of validity of the present model. 
%


\subsection{Dissipated and absorbed power by the oscillators}
We focus now on the dissipated and absorbed power by the shuttles.
According to  Eq. (\ref{eq:dyn}), the (unitless) power loss per unit cycle of duration $T$ will be given by
\[
\langle W_{\mathrm{dis}}\rangle=  \gamma{\langle \dot x^2\rangle}  = \frac{1}{2}\gamma{r_0^2}. 
\]
Similarly, we can obtain the absorbed power per unit cycle
by averaging over a period the pumped energy of the electrostatic force
given by the last term of Eq. (\ref{eq:dyn},  
\[
\langle W_{\mathrm{a}}\rangle\simeq \left\langle {\alpha^\prime n(t)\dot x(t)(\sin{\omega t} + \beta)}\right\rangle.
\]
Due to the correlation between charge fluctuations $n(t)$ and the velocity
of the nanopillars, $\dot x(t)$, a positive amount of energy may be pumped into
the system. 
For instance, for the finite dc bias cases, 
the amount of energy pumped into the system per cycle in steady state [at ($A_{\rm eq}$, $B_{\rm{eq}}$)$_p$] is  
\[
\langle W_{\mathrm{a}}\rangle =
\frac{r_0 
\alpha^\prime}{12} \left[6\beta n_0 \cos{\varphi} + 
(6n_{\mathrm{av}}\delta_{p,1} + 2{n_{0}}\delta_{p,2}- n_{0}\delta_{p,4})\sin{\varphi}\right]. 
\]
If this amount is larger than the dissipated power,
$\langle W_{\mathrm{a}}\rangle\gtrsim \langle W_{\mathrm{dis}}\rangle$,
self-sustained oscillations 
are expected. 
This may occur
when max$\{\alpha^\prime n_{\mathrm{av},0}, \alpha^\prime n_0\beta\} \gtrsim \gamma r_0$,
with the appropriate phase $\varphi$. 
In particular, for the branch with $\varphi \sim 0$, the condition 
reads $\alpha^\prime \beta n_0 > \gamma r_0 $. 
In other words, self-sustained oscillations may occur for a wide range of 
frequencies if the appropriate values of $\alpha$ and $\beta$ are met. 

\section{Conclusions}
\label{conclusions}
We have theoretically studied a coupled shuttle consisting of two oscillating 
nanoislands connected in series between two contacts. 
We express the chemical potentials in terms of the relative distance 
of the islands, $\mu_i^\leftrightarrows \left[x(t)\right]$,  and numerically
integrate the master equation to obtain the direct current through the system. 
Under a dc bias, Coulomb blockade diamonds were obtained. 
Adding  an rf signal, we analyze the response within the context of nonlinear dynamics. 
We study qualitatively and quantitatively the structure of the mode-locked tongues in 
the parameter space. 
Parametric instabilities are observed in 
a range of applied voltages and frequencies, where resulting small mechanical 
oscillations are amplified. 
In this instability region, an rf signal can be exploited to 
parametrically amplify the response to a gate excitation.
Hence, we propose a practical scheme for direct detection of instabilities in the 
mechanical motion of nanoscale objects. 

\begin{acknowledgments}
We are grateful to R. H. Blick and C. Kim for enlightening discussions.
This work was supported by the Spanish Ministry of Education, program SB2009-0071.
\end{acknowledgments}
\newpage
\appendix
\section{Free energy of a double coupled metallic island}
\label{app:fe}
In this appendix we derive the free energy of a double metallic grain system. 
In general, for a system of $N$ conductors, the total charge on each 
node $j$ is the sum of the charges on all of the capacitors connected to node $j$, 
$-en_j = -e\sum_k m_{k} = \sum_k C_k (V_j-V_k)$, where $V_j$ is the electrostatic 
potential of node $j$ and ground is defined to be at zero potential. The charges on 
the nodes are linear functions of the potential of the nodes, $\vec{Q} = \tilde{C}\vec{V}$, 
where $\tilde{C}$ is the capacitance matrix.
\begin{figure}[!hbt]
\centering\includegraphics[angle=0, width = 0.350\textwidth]{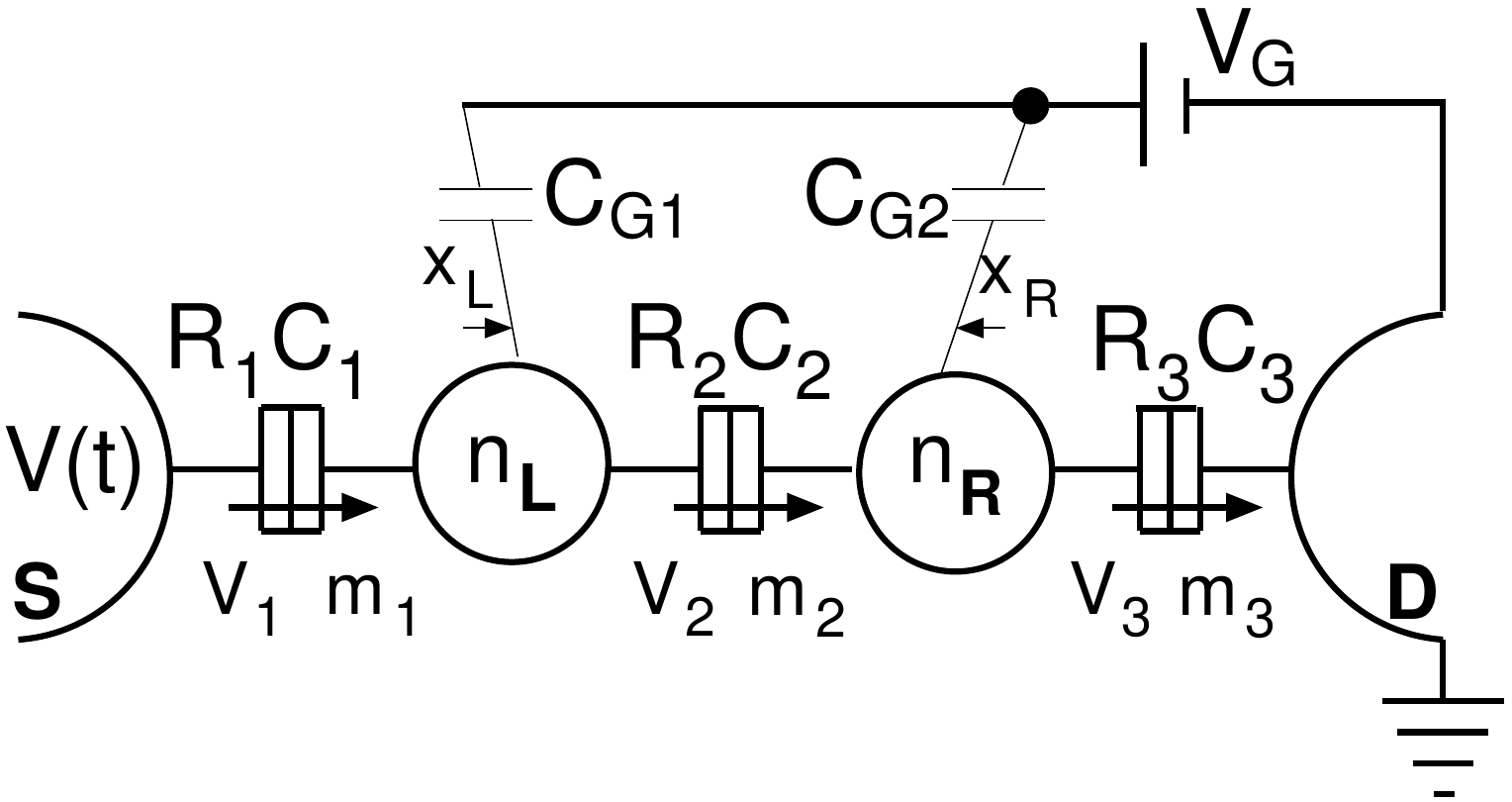}
\caption{\it \footnotesize A schematic picture of the double-island structure
with the voltage sources and various capacities in the system.
The three junctions are characterized by $m_j$ and $V_j$, $j$ = 1, 2, 3. 
The islands are coupled to each other, with a mutual capacitance $C_2$, 
as well as to a gate voltage, $V_G$ with capacitances $C_{G1}$ and 
$C_{G2}$, respectively. Also, they are coupled to the source-drain
leads, with capacitances $C_1$ and $C_3$. 
}
\label{figapp1}
\end{figure}
For the system depicted in Fig. \ref{figapp1}, we have:
\begin{eqnarray}
\label{eqQLR}
Q_L &=& 
C_{2}(V_{L}-V_R) + C_1(V_L-V_S) + C_{G1}(V_L- V_{G}) \nonumber \\
Q_R &=& 
C_2 (V_R-V_L)  + C_{3}(V_R-V_{D}) +C_{G2}(V_R-V_{G}), 
\end{eqnarray}
where $Q_i^{\rm{bg}}$ is the ``residual'' charge in dot $i$ when all potentials are grounded. 
We can write this in the form $\vec{Q} + \vec{\alpha} = \tilde{C} \vec{V}$, with $\tilde{C}$ 
being the capacitance matrix,
\[
\left(\begin{array} {c} Q_L+ C_1V_S + C_{G1}V_G \\ Q_R + C_3V_D+C_{G2}V_G
\end{array}\right) = 
\left(\begin{array} {c c} C_L & -C_2 \\ -C_2 & C_R  
\end{array}\right)
\left(\begin{array} {c} V_L \\ V_R 
\end{array}\right), 
 \]
where $C_L = C_1 + C_2+ C_{G1}$ and $C_R = C_3 + C_2+ C_{G2}$. 
We can thus set our node voltages in terms of the capacitances, 
\[
\left(\begin{array} {c} V_L \\ V_R 
\end{array}\right) = \frac{1}{C_LC_R - C_2^2}  
\left(\begin{array} {c c} C_R & C_2 \\ C_2 & C_L 
\end{array}\right)
\left(\begin{array} {c} Q_L+ C_1V_S + C_{G1}V_G \\ Q_R + C_3V_D+C_{G2}V_G
\end{array}\right).  
\]
The electrostatic energy of the system can now be calculated, $U = \vec{V} \tilde{C}\vec{V}/2$, 
with $V_S=V_D = V_G = 0$,
\[
U(n_L,n_R) = \frac{e^2}{C_LC_R-C_2^2}\left[
\frac{1}{2}C_Rn_L^2 + \frac{1}{2}C_L n_R^2 + C_2 n_Ln_R 
\right]. 
\]
To calculate the free energy, we first calculate the work performed by the external sources to achieve a configuration 
with $n_L = m_1-m_2$ and $n_R = m_2 - m_3$ electrons in the system. 
We consider Eq. (\ref{eqQLR}) in terms of the voltages on the junctions, $V_j$, $j$ = 1, 2, 3. 
For simplicity, we consider, first, the case without $V_G$, {\it i.e.}, $C_{G1/2}$ = 0: 
\begin{eqnarray}
V_1 & = & \frac{1}{\Sigma}
\left[ C_2C_3V(t) + em_1(C_2+C_3) - em_2C_3-em_3C_2\right]
\nonumber \\
V_2 & = & \frac{1}{\Sigma}
\left[ C_1C_3V(t) - em_1C_3 - em_2(C_3+C_1)-em_3C_1\right]
\nonumber \\
V_3 & = & \frac{1}{\Sigma}
\left[ C_1C_2V(t) - em_1C_2 - em_2C_1+em_3(C_1+C_2)\right],
\nonumber \\
\label{eq:vis}
\end{eqnarray}
where we have defined $\Sigma = C_LC_R-C_2^2$. We now 
calculate the work done by the external source, $V(t)= V_S$. 
For instance, the work done by $V_S$ for one electron to tunnel through the first 
junction, $m_1\to m_1+1$, is given by $\delta W_1 = V_S \delta{q_1}$, where 
$\delta{q_1} = -|e| + C_1\delta V_1$. From Eq. (\ref{eq:vis}), we see that 
\[
\delta V_1 = |e| \frac{C_2+C_3}{\Sigma} \to \delta W_1 = -|e|V_S \frac{C_2C_3}{\Sigma}. 
\]
In general, the work performed by the external source $V_S$ for a tunneling event
through the $i$-th junction is
\[
\delta W_i = -|e| V_S \frac{(\epsilon_{ijk})^2 C_jC_k }{2\Sigma} , 
\]
where $\epsilon_{ijk}$ is the Levi-Civita antisymmetric tensor. 
For the finite $C_{G1/2}$ case, we have (now $C_L = C_1 + C_2 + C_{G1}$ and 
$C_R = C_2 + C_3 + C_{G2}$):  
\begin{eqnarray}
V_1 & = & \frac{1}{\sigma}\left[ 
C_R(C_L -C_1) V(t) - (C_RC_{G1} + C_2C_{G2})V_G + \right.\nonumber \\ 
&+& \left.|e| (m_1C_R - m_2(C_R-C_2) - m_3C_2) \right]. \nonumber
\end{eqnarray}
We can now write an expression for $\delta W_i$ in terms of the capacitances, 
\begin{eqnarray}
\delta W_1& =& -|e| V_S \frac{C_2C_3 +C_{G1}C_R + C_{G2}C_2}{\Sigma}; \nonumber\\
\delta W_2& =& -|e| V_S \frac{C_1C_3+C_1C_{G2}}{\Sigma}; \nonumber \\ 
\delta W_3& =& -|e| V_S \frac{C_1C_2}{\Sigma}. \nonumber  
\end{eqnarray}
Likewise, the work performed by the gate in the tunneling process
through the $i$-th junction, 
$\delta W_{Gi} = V_G[\delta q_{G1}(m_i\to m_i+1)+ \delta q_{G2}(m_i\to m_i+1)]$: 
\begin{eqnarray}
\delta W_{G1}& =& |e| V_G \frac{C_{G1}C_R +C_{G2}C_2}{\Sigma}; \nonumber\\
\delta W_{G2}& =& |e| V_G \frac{C_{G2}(C_L-C_2) -C_{G1}(C_R-C_2)}{\Sigma}; \nonumber\\
\delta W_{G3}& =& -|e| V_G \frac{C_{G2}C_L+C_2C_{G1}}{\Sigma}. \nonumber \\ 
\end{eqnarray}
The electrostatic free energy is then given by $F = U - W$, with 
$W(\{m_i\}) = \sum m_i (\delta W_i + \delta W_{Gi})$, 
and, bearing in mind Eq. (\ref{eq:ns}),
\begin{widetext}
\begin{eqnarray}
F(\{n_\alpha,m_i\}) &=& \frac{e^2}{C_{L}C_{R}-C_2^2}  \left\{
C_R\frac{n_L^2}{2} + C_L\frac{n_R^2}{2} + C_2n_Ln_R  
- \frac{m_1}{|e|} \left[
V_G(C_{G1}C_R+ C_{G2}C_2 ) - V_S [C_2C_3 +C_{G1}C_R +C_{G2}C_2 ] 
\right] 
\right.\nonumber \\ &-& \left.
 \frac{m_2}{|e|}\left[
V_G(C_{G1}(C_R-C_2) -C_{G2}(C_L-C_2) ) + V_SC_1C_3+C_1C_{G2}  \right]
+ \frac{m_3}{|e|} \left[ 
V_G ( C_L C_{G2}+ C_2C_{G1}) +V_S {C_1C_2}  \right]
\right\}.\nonumber \\
\end{eqnarray}
\end{widetext}

\section{Transient equations} 
\subsection{rf excitation}
\label{app1}
We want to arrive from Eq. (\ref{eqx1}) to  Eq. (\ref{eqAB}).
We use $x_0 = A(\eta)\cos{z} + B(\eta)\sin{z}$ and Eq. (\ref{eqexx}):
\begin{eqnarray}
\frac{\partial x_0}{\partial z} &=& -A\sin{z}+B\cos{z}; \quad
\frac{\partial^2x_0}{\partial z^2} = -A\cos{z}-B\sin{z}; 
\nonumber \\ 
\frac{\partial^2x_0}{\partial z\partial \eta} &=& -\frac{\partial A}{\partial \eta}\sin{z}+
\frac{\partial B}{\partial \eta}\cos{z}. \nonumber
\label{eq:ders}
\end{eqnarray}
Also, we need the terms in $x_0$, $x_0^2$, and $x_0^3$ all multiplied by $(1-\cos{2z})$.
For the linear terms, we have to evaluate $(1-\cos{2z})x_0$, 
for the quadratic terms, $(1-\cos{2z})x_0^2$,
and, finally, for the cubic ones, $(1-\cos{2z})x_0^3$,
\begin{widetext}
\begin{eqnarray}
(1-\cos{2z})(A\sin{z}+B\cos{z}) &=& \frac{A}{2}(\cos{z} - \cos{3z}) + \frac{B}{2}(3\sin{z}-\sin{3z}) \nonumber \\
(1-\cos{2z})(A\sin{z}+B\cos{z})^2 &=& \frac{A}{4}(1-\cos{4z}) + 
\frac{AB}{2}(2\sin{2z}-\sin{4z})+
\frac{B^2}{4}(3-2\cos{2z} +\cos{4z}) \nonumber \\
(1-\cos{2z})(A\sin{z}+B\cos{z})^3 &= &
\frac{A^3}{4}\left(\cos{z} -\frac{1}{2}\cos{3z} -\frac{1}{2}\cos{5z} \right) +
\frac{3A^2B}{4}\left(\sin{z} +\frac{1}{2}\sin{3z} -\frac{1}{2}\sin{5z} \right) \nonumber \\ 
&+& \frac{3AB^2}{4}\left(\cos{z} -\frac{3}{2}\cos{3z} +\frac{1}{2}\cos{5z} \right) +
\frac{B^3}{4}\left(5\sin{z} -\frac{5}{2}\sin{3z} +\frac{1}{2}\sin{5z} \right) .
\nonumber
\end{eqnarray}
\end{widetext}
Now all we need to do is to substitute the above equations into  Eq. (\ref{eqx1}) and
arrange terms, {\it i.e.}, we obtain an equation of the form:
\[
\frac{\partial^2x_1}{\partial z^2}+ x_1 = (\dots)\sin{z} + (\dots)\cos{z} + {\mathrm{nonresonant}}\ {\mathrm{terms}}. 
\]
For non-resonant terms, we require the coefficients of $\sin{z}$ and $\cos{z}$ to vanish.
We then get Eq. (\ref{eqAB}).
\subsection{Superimposed DC excitation}
\label{app2}
We consider Eq. (\ref{eqnldyn3}) and as before, expand the solutions  in terms of
$\epsilon$, using Eqs. (\ref{eqexx}), (\ref{eqexw}), and (\ref{eqsol})
and arrive at a similar set of equations, only now we have extra terms 
(in $\beta, \beta^2$),
\begin{eqnarray}
&&\frac{\ud^2x_1}{\ud z^2} + x_1  + 
2\delta_\omega\frac{\ud^2x_0}{\ud z^2} + 2\frac{\ud^2x_0}{\ud z\ud \eta} +
\frac{1}{Q}\frac{\ud x_0}{\ud z} + \nonumber \\ 
&+& 2\alpha(\sin^2{z} + 2\beta\sin{z}+\beta^2)
\left(x_0-\frac{2\lambda}{d}x_0^2 - \frac{3}{16}x_0^3\right) = 0
\nonumber\\
\label{eqna}
\end{eqnarray}
The last term is a tedious one, giving nine terms, six of which are new. We use 
\begin{eqnarray}
2\alpha \sin^2{z}x_0 &\simeq &\frac{\alpha}{2}[A\cos{z} + 3 B\sin{z}] \nonumber\\
-\frac{4\lambda\alpha}{d} \sin^2{z}x^2_0 &\simeq &-\frac{\lambda\alpha}{2d}[A^2 + 3 B^2] \nonumber\\
-\frac{6\alpha}{16} \sin^2{z}x^3_0 &\simeq &-\frac{3\alpha}{64}
[A(A^2+3B^2)\cos{z} +\nonumber \\&&+ B(3A^2+5B^2)\sin{z}]\nonumber\\
4\alpha\beta\sin{z}x_0&\simeq &2\alpha\beta B\nonumber\\
-\frac{8\lambda\alpha\beta}{d}\sin{z}x^2_0&\simeq & 
-\frac{ 2\lambda\alpha\beta}{d}B[(A^2+3B^2)\sin{z}+2AB\cos{z}]
\nonumber\\
-\frac{12\alpha}{16} \sin{z}x^3_0 &\simeq &-\frac{9\alpha\beta}{32}B(A^2+B^2)\nonumber\\
2\alpha\beta^2x_0&\simeq &2\alpha\beta^2 ( A\cos{z} + B\sin{z} )\nonumber\\
-\frac{4\lambda\alpha\beta^2}{d}x_0^2&\simeq & -\frac{2\lambda\alpha\beta^2}{d}(A^2+B^2)  \nonumber\\
-\frac{3\alpha\beta^2}{16}x_0^3&\simeq & -\frac{9\alpha\beta^2}{32}(A^2+B^2)
(A\cos{z} + B\sin{z}) . \nonumber
\end{eqnarray}
Substituting into Eq. (\ref{eqna}) and again arranging terms in $\sin{z}$ and $\cos{z}$,
\begin{eqnarray}
2\frac{\partial A }{\partial \eta} &= &
-\gamma A -\left(2\delta_\omega - \frac{3\alpha}{2} -2 \alpha\beta^2\right)B- 
\frac{2\lambda\alpha\beta}{d}(A^2+3B^2) -\nonumber \\ &&-
\frac{3\alpha}{64}B[3A^2(1+2\beta^2) + B^2(5+6\beta^2) ] 
\nonumber \\
2 \frac{\partial B }{\partial \eta} &= &
-\gamma B +\left(2\delta_\omega - \frac{\alpha}{2} -2\alpha\beta^2\right)A+ 
\frac{4\lambda\alpha\beta}{d}AB +\nonumber \\ &&+
\frac{3\alpha}{64}A[A^2(1+6\beta^2) + 3B^2(1+2\beta^2)]
\label{eqAB3}
\end{eqnarray}
\subsection{Subharmonics in the general case}
\label{app1b}
We consider the general case in which the driving frequency is given by
$ \omega \simeq p + \epsilon \delta_\omega$ and $x\simeq x_0 + \epsilon x_1$, giving: 
\begin{eqnarray}
\dot x(t) &=& 
           p\frac{\ud x_0}{\ud z} + \epsilon \delta_\omega \frac{\ud x_0}{\ud z} + 
         \epsilon \frac{\ud x_0}{\ud \eta} + \epsilon p \frac{\ud x_0}{\ud z} + O(\epsilon^2),
\nonumber \\
\ddot x(t) &=&
          p^2\frac{\ud^2 x_0}{\ud z^2} + 2 \epsilon \delta_\omega p \frac{\ud^2 x_0}{\ud z^2} + 
         2\epsilon p \frac{\ud^2 x_0}{\ud \eta\ud z} + \epsilon p^2 \frac{\ud^2 x_1}{\ud z^2} + O(\epsilon^2).
\nonumber \\
\label{eqxders}
\end{eqnarray}
So now Eq. (\ref{eqnldyn3}) is, to lowest order in $\epsilon$, 
\[
p^2\frac{\ud^2 x_0}{\ud z^2} + x_0 = 0 \quad \to \quad x_0 = A(\eta) \cos{\frac{z}{p}} + B(\eta) \sin{\frac{z}{p}}. 
\]
Substituting the expression for $x_0$ into Eq. (\ref{eqnldyn3}) to first order in $\epsilon$, we get 
\begin{widetext}
\begin{eqnarray}
 &-&\frac{2\delta_\omega}{p} \left[A(\eta) \cos{\frac{z}{p}} + B(\eta) \sin{\frac{z}{p}}\right] 
- 2\left[ \frac{\ud A(\eta)}{\ud \eta} \sin{\frac{z}{p}} - \frac{\ud B(\eta)}{\ud \eta} \cos{\frac{z}{p}}\right] 
- \frac{1}{Q} \left[A(\eta) \sin{\frac{z}{p}} + B(\eta) \cos{\frac{z}{p}}\right]  
\nonumber \\
&+& 2\alpha(\sin^2{z} + 2\beta\sin{z}+\beta^2)
\left(
\left[A(\eta) \cos{\frac{z}{p}} + B(\eta) \sin{\frac{z}{p}}\right]
-\frac{2\lambda}{d}\left[A(\eta) \cos{\frac{z}{p}} + B(\eta) \sin{\frac{z}{p}}\right]^2 
\right. \nonumber \\ &-&\left. 
 \frac{3}{16}\left[A(\eta) \cos{\frac{z}{p}} + B(\eta) \sin{\frac{z}{p}}\right]^3
\right) = 0
\label{eqp}
\end{eqnarray}
\end{widetext}
Next, we need to evaluate the last nine terms of the equation above, which involves
trigonometric operations. 
The relevant contributions for the dynamics of the system are the terms
proportional to $\sin{z/p}$ and $\cos{z/p}$ (resonant terms). 
These are: 
\[
\alpha x_0\sin^2z 
= \frac{\alpha}{4} \left[A(2+\delta_{p,1}) \cos{\frac{z}{p}} + B (2-\delta_{p,1})\sin{\frac{z}{p}} \right]  
\]
\[
2\beta\alpha x_0 \sin{z} = \beta\alpha \left[
A\sin{\frac{z}{p}}+B\cos{\frac{z}{p}}
\right]\delta_{p,2}
\]
\[
\beta\alpha^2 x_0  = \beta\alpha^2 \left[
A\sin{\frac{z}{p}}+B\cos{\frac{z}{p}}
\right]
\]
\[
\frac{\lambda\alpha x^2_0}{2d}\sin^2z = \frac{\lambda\alpha}{16d}  \left[(B^2-A^2)\cos{\frac{z}{p}} +AB \sin{\frac{z}{p}} \right]\delta_{p,3/2}
\]
\[
\frac{\lambda\alpha\beta}{d} x_0^2 \sin{z} = 
\frac{\lambda\alpha\beta}{4d}\left[
2AB\cos{\frac{z}{p}}+(A^2-B^2)\sin{\frac{z}{p}}
\right]\delta_{p,3}
\]
\[
\frac{\lambda\alpha\beta^2}{2d} x_0^2 \sin{z} \to 
{\mathrm{no}}\ {\mathrm{contributions}}
\]
\[
-\frac{3\alpha\beta^2 x^3_0}{16} = 
-\frac{9\alpha\beta^2}{64}(A^2+B^2) 
\left[
A\sin{\frac{z}{p}}+B\cos{\frac{z}{p}}
\right]
\]
\begin{widetext}
\begin{eqnarray}
-\frac{3\alpha x^3_0}{16}\sin^2z& = & 
-\frac{3\alpha}{128}  \left[
A\cos{\frac{z}{p}}\left(3(A^2+B^2)(1 - \frac{3}{2}\delta_{p,1})-\frac{1}{2}(3A^2-B^2)\delta_{p,2}  \right)  
\right.\nonumber\\ && \left.
+B \sin{\frac{z}{p}} \left(3(A^2+B^2)(1 - \frac{3}{2}\delta_{p,1})-\frac{1}{2}(3A^2-B^2)\delta_{p,2}  \right)
\right]
\nonumber
\end{eqnarray}
\begin{eqnarray}
-\frac{3\alpha x^3_0}{16}\sin^2z& =  & 
-\frac{3\alpha}{128}  \left[
A\sin{\frac{z}{p}} \left(3(A^2+B^2)\delta_{p,2} + (A^2-3B^2)\delta_{p,4}  \right)  
\right.\nonumber\\ && \left.
+B \cos{\frac{z}{p}} \left(3(A^2+B^2)\delta_{p,2}+(3A^2-B^2)\delta_{p,4}  \right)
\right]
\nonumber
\end{eqnarray}
\end{widetext}
As we did before, we can now arrange all the terms in Eq. (\ref{eqp})
as coefficients of $\sin{z/p}$ and $\cos{z/p}$, resulting in our desired equation,
\begin{widetext}
\begin{eqnarray}
2\frac{\partial A }{\partial \eta} &= &
-A\left(\gamma-\beta\alpha\delta_{p,2}\right) - 
B \left(\frac{2\delta_\omega}{p} - \frac{\alpha}{4}(2+  \delta_{p,1}+4\beta^2)\right)+ 
AB\frac{\lambda\alpha}{8d}\delta_{p,3/2} + (A^2-B^2)\frac{\beta\lambda\alpha}{4d}\delta_{p,3}
\nonumber \\ && 
-\frac{3\alpha}{256}\left[ B(A^2+B^2)(6+12\beta^2 +3\delta_{p,1}) + 
   B(3A^2-B^2)\delta_{p,2} +6A(A^2+B^2)\delta_{p,2} + 2A(A^2-3B^2)\delta_{p,4}  \right] 
\nonumber \\
2 \frac{\partial B }{\partial \eta} &= &
-B\left(\gamma+\beta\alpha\delta_{p,2}\right) + 
A \left(\frac{2\delta_\omega}{p} - \frac{\alpha}{4}(2-  \delta_{p,1}+4\beta^2)\right)+ 
(A^2-B^2)\frac{\lambda\alpha}{16d}\delta_{p,3/2}  
- AB\frac{\lambda\beta\alpha}{2d}\delta_{p,3}
\nonumber \\ && 
+\frac{3\alpha}{256}\left[ A(A^2+B^2)(6+12\beta^2 -3\delta_{p,1}) - 
   A(A^2-3B^2)\delta_{p,2} +6B(A^2+B^2)\delta_{p,2} + 2B(3A^2-B^2)\delta_{p,4}  \right] 
\nonumber\\
\label{eqAB4}
\end{eqnarray}
\end{widetext}
\subsection{Transients in the Coulomb blockade limit}
\label{app:CB}
We substitute again the general 
solution, $x(t) \simeq x_0 + \epsilon x_1$, 
in Eq. (\ref{eq:dyn}), along with the expression for $\langle n(t) \rangle$ 
given in (\ref{eq:n}). 
The equation of motion now reads  
\begin{eqnarray}
\frac{\partial^2x_1}{\partial z^2} + x_1  
+2\frac{\partial^2x_0}{\partial z\partial \eta} 
+2\delta_\omega\frac{\partial^2x_0}{\partial z^2} -\frac{1}{Q_1} \frac{\partial x_0}{\partial z} + &&
\nonumber \\ 
\alpha_1^\prime(\sin{\omega t} +\beta) (n_{\mathrm{av}} + 4n_0(\cos{\omega_0t} - \cos{3\omega_0t})/\pi) &=& 0
\nonumber \\
\label{eqx2}
\end{eqnarray}
As before, we evaluate the equation of motion with $x_0(t) \simeq A(\eta) \cos{z/p} + B(\eta)\sin{z/p}$, 
using  (\ref{eq:ders}). 
Inserting this into Eq. (\ref{eqx2}) and considering 
the resonant terms, with $\omega t = z$ and $\omega_0 t = z/p$, we obtain  
\begin{widetext}
\begin{eqnarray}
-2\frac{\delta_\omega}{p}A \cos{\frac{z}{p}} 
-2\frac{\delta_\omega}{p}B \sin{\frac{z}{p}} 
-2\frac{\ud A}{\ud\eta} \sin{\frac{z}{p}} 
+2\frac{\ud B}{\ud\eta} \cos{\frac{z}{p}} 
-\gamma_1 A \sin{\frac{z}{p}} 
+\gamma_1 B \cos{\frac{z}{p}} 
+ \alpha_1^\prime \beta n_{\mathrm{av}}
&&\nonumber \\
+ \alpha_1^\prime \beta n_{0}\cos{\frac{z}{p}} 
+ \alpha_1^\prime  n_{\mathrm{av}}\sin{\frac{z}{p}}\delta_{p,1}
- \frac{1}{2}\alpha_1^\prime n_{\mathrm{0}}\sin{\frac{z}{p}}\delta_{p,2}
-\frac{1}{6}\alpha_1^\prime n_{\mathrm{0}}\sin{\frac{z}{p}}(\delta_{p,4}-\delta_{p,2}) 
&=&\ddot x_1 + x_1.
\end{eqnarray}
\end{widetext}
Arranging the terms in $\sin{z/p}$ and $\cos{z/p}$, we get Eq. (\ref{eq:AB_CB}).

In the small oscillations limit, $n(t)$ ``follows'' linearly the mechanical oscillations of the 
islands, $n(t) \simeq n_0 \sin{\omega_0 t}$, giving a different set of equations, 
\begin{eqnarray}
\label{eq:CBlin}
2\frac{\ud A }{\ud \eta} &= &
-\gamma_1 A -  \frac{2\delta_\omega}{p}  B
+ \alpha_1^\prime\beta n_0, 
\nonumber \\
2 \frac{\ud B }{\ud \eta} &= &
-\gamma_1 B +  \frac{2\delta_\omega}{p} A
- n_0\frac{\alpha^\prime_1}{3}\delta_{p,2} .
\end{eqnarray}

\end{document}